\documentclass[twocolumn,twoside]{IEEEtran}

\usepackage{amssymb,amsmath,amsfonts}
\usepackage{color}
\usepackage{mathtools}
\usepackage{mathrsfs}
\usepackage{bm}
\usepackage{graphicx}	
\usepackage[font={small}]{caption}
\usepackage{caption}
\usepackage{subcaption}
\usepackage{multicol}
\usepackage{enumerate}
\usepackage{units}
\usepackage[numbers]{natbib}
\usepackage{algorithm}
\usepackage{color}

\usepackage{cuted}

\DeclarePairedDelimiter\abs{\lvert}{\rvert}

\begin{document}

\title{Spectral Efficiency of OFDMA Cognitive Radios under Imperfect Cross-Link Knowledge}

\author{\IEEEauthorblockN{H. Saki and M. Shikh-Bahaei}
\IEEEauthorblockA{Institute of Telecommunications, King's College London, London, WC2R 2LS, United Kingdom\\
Email: \{hadi.saki, m.sbahaei\}@kcl.ac.uk}}

%
%

\maketitle

\begin{abstract}
We analyse the spectral efficiency performance and limits of orthogonal frequency-division multiple access (OFDMA) cognitive radios (CRs) with imperfect availability of cross-link knowledge. In particular, in contrast to the conventional `average' and `worst' cases of channel estimation error in the literature, we propose a stochastic approach to mitigate the total imposed interference on primary users. Channel-adaptive resource allocation algorithms are incorporated to optimize the cognitive system functionality under transmit and interference power constraints. An expression for the cumulative density function (cdf) of the received signal-to-interference-plus-noise ratio (SINR) is developed to evaluate the average spectral efficiency. Analytical derivations and results are confirmed through computer simulations. 

\end{abstract}

\section{Introduction}
Dynamic spectrum management (DSM) has attracted a lot of attention recently as measurements by spectrum regulators have shown that the radio frequency (RF) band is severely underutilized.

According to the Office of Communications (Ofcom) recent measurements \cite{7678715}, the spectrum is severely under-utilized as a result of rigid and inefficient management policies, and that 90\% of locations have around 100 MHz of spectrum available for other services. Flexible spectrum-sharing for supporting CR and spectrum co-existence is a priority issue to overcome the current capacity crunch and thus enabling the deployment of long term evolution (LTE)-advanced and beyond wireless systems. 

In recent years, a significant effort has been made towards improving the spectral efficiency of cellular networks in order to meet the growing demand and sophistication of wireless applications. Several spectral-efficient technologies, such as cross layed design \citep{ShojaeifardZS11}, machine-to-machine (M2M) communications, small-cell (SC) solution, massive multiple-input multiple-output (MIMO), and cognitive radio (CR) - each with respective advantages and challenges - are promising candidates in this direction \cite{6191306}, \citep{kob}. The term CR can be defined as an intelligent radio system that has the ability to sense the primary service behaviour and surrounding environment and adjust its spectrum usage and parameters based on the observed information \cite{788210}, \cite{SakiMS16}. 

Three main paradigms have been proposed for cognitive radio in regards to the unlicensed users access to the primary frequency band: (i) underlay spectrum access where secondary users silently coexist with primary users, provided they satisfy an interference limit set by a regulatory authority (ii) overlay spectrum access in which secondary users are only allowed to access the vacant parts of the primary spectrum, and (iii) hybrid spectrum access, a combination of the two former strategies in which the secondary users sense the primary spectrum and adjust their transmission parameters based on the detection, whilst avoid imposing harmful interference to the primary users \cite{4481339}. In this paper, we consider underlay spectrum-sharing, where robust interference management is critical for tackling any harmful cross-service interference. 

Orthogonal frequency-division multiplexing (OFDM) has emerged as a prominent radio access technology for new generation of wireless communication systems including LTE and LTE-advanced \cite{995857},\cite{SakiS15}. Unilike CDMA \cite{ZarringhalamSSCA09}, OFDM-based multi-user applications, multiple-access can be accommodated through orthogonal frequency-division multiple-access (OFDMA) technique \cite{4027580}. In OFDMA systems, different subcarriers may be assigned to different users in order to exploit the channel quality random variations of users across each subcarrier. OFDMA technology is considered as a de facto standard for CR networks due to its inherent advantages in terms of flexibility and adaptability in allocating spectrum resources in shared-spectrum environments \cite{5510778}. 

Radio resource allocation (RRA) plays a significant role in optimizing the overall spectral efficiency of conventional OFDMA systems \cite{4389761}. In addition, adaptive RRA is an active area of research in the context of OFDMA-based CR networks with the aim of achieving a balance between maximizing the cognitive network performance and minimizing the inflicted interference on the licensed users. Suboptimal and optimal power allocation policies are studied in \cite{4686852}, where the aggregate capacity of the CR system is maximized under a primary receiver (PRx) interference limit. In \cite{5557657}, a queue-aware RRA algorithm is proposed to maximize the fairness in OFDMA-based CR networks subject to a total power constraint at the base station. A Lagrangian relaxation algorithm is adopted in \cite{5510778} to probabilistically allocate resources based on the availability of the primary frequency band via spectrum sensing. 

Most of the RRA algorithms on CR networks in the literature assume perfect channel state information (CSI) between the cognitive transmitter (CTx) and PRx, and few have considered imperfect cross-link CSI. However, due to technical reasons such as estimation errors and wireless channel delay, obtaining perfect cross-link CSI is difficult in practical scenarios.
In \cite{5419086} and \cite{6185693}, the ergodic capacity is derived over fading channels with imperfect cross-link knowledge, however, the analysis is carried out for a single cognitive user (CU). Furthermore, due to noisy cross-link information, it is unrealistic to assume that the secondary network strictly satisfies a deterministic interference constraint. The authors in \cite{5967979} propose a RRA algorithm for maximizing instantaneous rate in downlink OFDMA CR systems subject to satisfying a collision probability constraint. However, \cite{5967979} only considers the individual impact of probabilistic interference constraint per subcarrier. Motivated by the above, we thoroughly investigate different scenarios by analysing the impact of deterministic and probabilistic interference constraints depending on perfect and noisy cross-link knowledge. In particular, we develop novel RRA algorithms under `average case', `worst case', and `probabilistic case' scenarios of channel estimation uncertainty for \textit{multi-user} OFDMA CR networks. 

On the other hand, to the best of authors' knowledge, enhancing the average spectral efficiency of multi-user OFDMA-based CR systems has not been addressed in the literature. In this work, by exploiting the advantages of channel adaptation techniques, we propose novel joint power, subcarrier, and rate allocation algorithms for enhancing the average spectral efficiency of downlink multi-user adaptive M-ary quadrature amplitude modulation (MQAM)/OFDMA \cite{1558992}, \citep{neh} CR systems. Given the received power restrictions on the CTx in order to satisfy the primary network interference limit and the cognitive network power constraint, the CTx transmit power is a function of the cognitive-cognitive direct-link and cognitive-primary cross-link fading states. We develop a cumulative distribution function (cdf) of the CR's received signal-to-interference-plus-noise ratio (SINR) to evaluate the average spectral efficiency of the adaptive MQAM/OFDMA CR system.  
      
The main novelties and contributions of this paper are summarized as follows:
\begin{enumerate}
\item The comprehensive problem of power, rate, and subcarrier allocation for enhancing the average spectral efficiency of downlink multi-user OFDMA CR systems subject to satisfying total average transmission power and peak aggregate interference constraint has been studied.

\item A closed-form expression for the cdf of the OFDMA CR's received SINR is derived under limitations imposed on the CTx through the power and interference constraints. Consequently, an upper-bound expression for average spectral efficiency of the adaptive multi-user MQAM/OFDMA CR system is formulated.

\item The critical issue of violating interference limits associated with imperfect cross-link CSI availability is examined by carrying out the analysis for the `average case', `worst case', and `probabilistic case' scenarios of channel estimation error. 

\item The impact of deterministic and probabilistic interference constraints on the system performance is considered with perfect and imperfect cross-link CSI. In particular, we propose a new low-complexity deterministic formulation for the probabilistic cross-link interference. 
\end{enumerate}

The organization of this paper is as follows: Section II presents the network model and operation assumptions. In Section III, the resource allocation problem for enhancing average spectral efficiency of the adaptive multi-user MQAM/OFDMA under perfect cross-link CSI subject to power and deterministic interference constraints is developed. In Section IV, under noisy cross-link knowledge, the impact of `average case' and `worst case' of channel estimation error based on a posterior distribution of the perfect channel conditioned on its estimate is examined. Section V investigates the performance under a collision probability constraint with imperfect cross-link CSI and proposes a deterministic formulation of the probabilistic aggregate cross-link interference. In all of the RRAs derived in the paper, optimal power, rate, and subcarrier assignments are obtained. Illustrative numerical results for various scenarios under consideration are provided in Section VI. Finally, concluding remarks are presented in Section VII. 

\section{System Model and Preliminaries}

In this section, the multi-user OFDMA CR network model, wireless channel, and operation assumptions are introduced. Further, interference management schemes and spectral efficiency of the adaptive MQAM/OFDMA system under consideration are studied.   

\subsection{Network Architecture and Wireless Channel}

\begin{figure}[t]
\centering
\includegraphics[width=.375\textwidth]{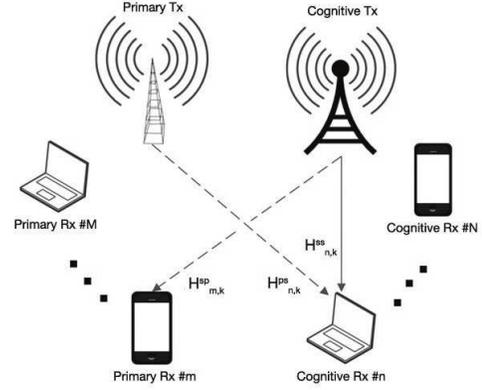}
\caption{Schematic diagram of the shared-spectrum OFDMA system. For simplicity purposes, channels of a single cognitive user are drawn.}
\label{CRmodel}
\end{figure}

We consider an underlay shared-spectrum environment, as shown in Fig. \ref{CRmodel}, where a cognitive network with a single CTx and $n = 1, ..., N$ cognitive receiver (CRx)s coexist with a primary network with a primary transmitter (PTx) and $m = 1, ..., M$ PRxs. The cognitive network can access a spectrum licensed to the primary network with a total bandwidth of $B$ which is divided into $K$ non-overlapping sub-channels subject to not violating the imposed interference constraint set by a regulatory authority. The sub-channel bandwidth is assumed to be much smaller than the coherence bandwidth of the wireless channel, thus, each subcarrier experiences frequency-flat fading. Let $H^{ss}_{n,k}(t)$, $H^{ps}_{n,k}(t)$, and $H^{sp}_{m,k}(t)$, at time $t$, denote the channel gains over subchannel $k$ from the CTx to $n^{\text{th}}$ CRx, PTx to $n^{\text{th}}$ CRx, and CTx to $m^{\text{th}}$ PRx. The channel power gains $|H^{ss}_{n,k}(t)|^2$, $|H^{ps}_{n,k}(t)|^2$, and $|H^{sp}_{m,k}(t)|^2$ are assumed to be ergodic and stationary with continuous probability density functions (pdf)s $f_{|H^{ss}_{n,k}(t)|^2}(.)$, $f_{|H^{ps}_{n,k}(t)|^2}(.)$, and $f_{|H^{sp}_{m,k}(t)|^2}(.)$, respectively. In addition, the instantaneous values and distribution information of secondary-secondary channel power gains is assumed to be available at the CTx \cite{6185693}. In this work, we consider different cases with perfect and noisy cross-link knowledge between CTx and PRxs. 

Each sub-channel is assigned exclusively to at most one CRx at any given time, hence, there is no mutual interference between different cognitive users \cite{6104418}. It should also be noted that by utilizing an appropriate cyclic prefix, the inter-symbol-interference (ICI) can be ignored \cite{5672622}. The received SINR of cognitive user $n$ over sub-channel $k$ at time interval $t$ is 
\begin{align}
\gamma_{n,k}(t) = \frac{P_{n,k} |H^{ss}_{n,k}(t)|^2}{\sigma^2_{n} + \sigma^2_{ps}}
\end{align}
where $P_{n,k}$ is a fixed transmit power allocated to cognitive user $n$ over sub-channel $k$, $\sigma^{2}_{n}$ is the noise power, and $\sigma^{2}_{ps}$ is the received power from the primary network. Without loss of generality, $\sigma^{2}_{n}$ and $\sigma^{2}_{ps}$ are assumed to be the same across all users and sub-channels \cite{5290301,5672619}. We define $\Upsilon_{n,k}(t)$ as a vector containing $\gamma_{n,k}(t)$ of all time intervals. For the sake of brevity, we henceforth omit the time reference $t$.    

Due to the impact of several factors, such as channel estimation error, feedback delay, and mobility, perfect cross-link information is not available. With noisy cross-link CTx to PRxs knowledge, we model the inherent uncertainty in channel estimation in the following form
\begin{align}
H^{sp}_{m,k} = \hat{H}^{sp}_{m,k} + \Delta H^{sp}_{m,k}
\label{errorF}
\end{align}
where over subcarrier $k$, $H^{sp}_{m,k}$ is the actual cross-link gain, $\hat{H}^{sp}_{m,k}$ is the channel estimation considered to be known, and $\Delta H^{sp}_{m,k}$ denotes the estimation error. $H^{sp}_{m,k}$, $\hat{H}^{sp}_{m,k}$, and $\Delta H^{sp}_{m,k}$ are assumed to be zero-mean complex Gaussian random variables with respective variances $\delta^{2}_{H^{sp}_{m,k}}$, $\delta^{2}_{\hat{H}^{sp}_{m,k}}$, and $\delta^{2}_{\Delta H^{sp}_{m,k}}$ \cite{5419086,4801449}. For robust receiver design, we consider the estimation $\hat{H}^{sp}_{m,k}$ and error $\Delta H^{sp}_{m,k}$ to be statistically correlated random variables with a correlation factor $\rho = \sqrt{\delta^{2}_{\Delta H^{sp}_{m,k}}/(\delta^{2}_{\Delta H^{sp}_{m,k}} + \delta^{2}_{H^{sp}_{m,k}})}$, where $0 \leq \rho \leq 1$. 

\subsection{Interference Management}

In a shared-spectrum environment, and particularly for delay-sensitive services, the licensed users' quality of service (QoS) is highly dependent to the instantaneous received SINRs of cognitive users. In order to protect the licensed spectrum from harmful interference we pose a deterministic peak total interference constraint between CTx and primary users
\begin{align}
\sum_{n=1}^{N} \sum_{k=1}^{K} \varphi_{n,k}(\Upsilon) P_{n,k}(\Upsilon) |H^{sp}_{m,k}|^2 \leq I^{m}_{th} , \forall m \in \{1, ..., M\}
\label{IMeq1}
\end{align}
where $\Upsilon$ is a matrix containing all of $\Upsilon_{n,k}$,  $\forall n \in \{1, ..., N\}$ and $\forall k \in \{1, ..., K\}$, further, $\varphi_{n,k}(\Upsilon)$ is the time-sharing factor (subcarrier allocation policy), $P_{n,k}(\Upsilon)$ is the allocated transmit power, and $I^{m}_{th}$ denotes the maximum tolerable interference threshold. 

However, as a consequence of uncertainties about the shared-spectrum environment and primary service operation, it is unrealistic to assume that the CTx always satisfies the deterministic peak total interference constraint. In practical scenarios, probability of violating the interference constraint is confined to a certain value that satisfies the minimum QoS requirements of primary users. Probabilistic interference constraint is particularly critical for robust interference management given noisy cross-link knowledge. To improve overall system performance and to mitigate the
impact of channel estimation errors, the following allowable probabilistic interference limit violation is considered
\begin{gather}
\mathscr{P} \left( \sum_{n=1}^{N} \sum_{k=1}^{K} \varphi_{n,k}(\Upsilon) P_{n,k}(\Upsilon) |H^{sp}_{m,k}|^2 > I^{m}_{th} \right) \leq \epsilon^{m} \nonumber \\ , \forall m \in \{1, ..., M\}
\label{IMeq2}
\end{gather}
where $\mathscr{P}(.)$ denotes probability, and $\epsilon^{m}$ is the collision probability constraint of $m^{\text{th}}$ PRx. 

On the other hand, mitigating the interference between neighbouring cells is a vital issue due to the increasing frequency reuse aggressiveness in modern wireless communication systems \cite{5770666}. As a remedy to inter-cell interference, and to maintain effective and efficient power consumption, we impose a total average transmit power constraint on the cognitive network as follows
\begin{align}
\sum_{n=1}^{N} \sum_{k=1}^{K} E_{\Upsilon} \Big\{ \varphi_{n,k}(\Upsilon) P_{n,k}(\Upsilon) \Big\} \leq P_t.
\end{align}
where $E_{x}(.)$ denotes the expectation with respect to $x$, and $P_{t}$ denotes the total average transmit power limit. 

\subsection{Spectral Efficiency}

The focus of this work is mainly on optimal power, rate, and subcarrier allocation for enhancing the average spectral efficiency of the adaptive MQAM/OFDMA CR network. In a multi-user scenario, various subcarriers may be allocated to different users. In other words, users may experience different channel fading conditions over each sub-channel. Therefore, any efficient resource allocation scheme in OFDMA must be based on the sub-channel quality of each user. Furthermore, in a shared-spectrum environment, satisfying the interference constraints is an important factor in allocating resources.   

Employing square MQAM with Gray-coded bit mapping, the approximate instantaneous bit-error-rate (BER) expression for user $n$ over subcarrier $k$ is given by
\begin{align}
\xi^{b}_{n,k}(\Upsilon) & = \frac{4}{\log_{2}(M_{n,k}(\Upsilon))} \Biggr(1 - \frac{1}{\sqrt{M_{n,k}(\Upsilon)}} \Biggr) \nonumber \\ & \times Q \Biggr(\sqrt{\frac{3 \Upsilon_{n,k}}{M_{n,k}(\Upsilon)-1}} \Biggr)
\label{approxBER}
\end{align} 
where $M_{n,k}(\Upsilon)$ denotes the constellation size vector of MQAM which each element is a function of the instantaneous received SINR of the cognitive user $n$ over subcarrier $k$, and $Q(.)$ represents the Gaussian Q-function. 

The aggregate average spectral efficiency of the adaptive multi-user MQAM/OFDMA system per subcarrier per user over the fading channel is defined as 
\begin{align}
ASE = \sum_{n=1}^{N} \sum_{k=1}^{K} E_{\Upsilon} \Big\{ \log_{2} \left( M_{n,k}(\Upsilon) \varphi_{n,k}(\Upsilon) \right) \Big\}.
\label{AASE}
\end{align} 
In order to evaluate the $ASE$, the distribution of the received SINR, a function of secondary-secondary and secondary-primary channels, must be developed. 

\section{Deterministic Interference Constraint with Perfect Cross-Link CSI}

The objective of this paper is to maximize the aggregate average spectral efficiency of cognitive users while satisfying total transmission power and peak maximum tolerable interference constraints. In this section, we solve the resource allocation problem with the perfect cross-link knowledge and deterministic interference constraint.

\subsection{Problem Formulation}
Mathematically, the optimization problem can be stated as follows.

\textit{Problem} $\mathscr{O}_{1}$:
\begin{subequations}
\label{optimizationProb}
\begin{gather}
\max_{\varphi_{n,k}(\Upsilon),P_{n,k}(\Upsilon)} \sum_{n=1}^{N} \sum_{k=1}^{K} E_{\Upsilon} \Big\{ \log_{2}(M_{n,k}(\Upsilon)) \varphi_{n,k}(\Upsilon) \Big\}  
\label{OF1a}\\
\text{s. t.:} \quad \sum_{n=1}^{N} \sum_{k=1}^{K} E_{\Upsilon} \Big\{  \varphi_{n,k}(\Upsilon) P_{n,k}(\Upsilon) \Big\} \leq P_t \label{OF1b}\\
\sum_{n=1}^{N} \sum_{k=1}^{K} \varphi_{n,k}(\Upsilon) P_{n,k}(\Upsilon) |H^{sp}_{m,k}|^2 \! \leq \! I^{m}_{th} , \forall m \! \in \! \{1, ..., M\} \label{OF1c}\\
\sum_{n=1}^{N} \varphi_{n,k}(\Upsilon) = 1 , \forall k \! \in \! \{1, ..., K\} \label{OF1d}\\
\varphi_{n,k}(\Upsilon) \in \{0,1\} , \forall n \in \{1, ..., N\} , \forall k \in \{1, ..., K\}  \label{OF1be}\\
\xi^{b}_{n,k}(\Upsilon) \leq \xi , \forall n \in \{1, ..., N\} , \forall k \in \{1, ..., K\} \label{OF1e}
\end{gather}
\end{subequations}
where $\xi$ denotes the common BER-target.

In the adaptive multi-user MQAM/OFDMA CR system under consideration, different transmit power and constellation sizes are allocated to different users and subcarriers. Using the upper-bound expression for the Gaussian Q-function, i.e., $Q(x) \leq (1/2) \exp(-x^2/2)$, the instantaneous BER for user $n$ over subcarrier $k$, subject to an instantaneous constraint $\xi^{b}_{n,k}(\Upsilon) = \xi$ can be expressed as 
\begin{align}
\xi^{b}_{n,k}(\Upsilon) \leq 0.3 \exp \Biggr( \frac{-1.5  \Upsilon_{n,k} }{M_{n,k}(\Upsilon) - 1 } \frac{P_{n,k}(\Upsilon)}{\min \left(\frac{P_t}{K} , \frac{I^{m}_{th}}{N^{sp}_{m}} \right)} \Biggr). 
\end{align}
where $N^{sp}_{m} = \sum_{k=1}^{K} |H^{sp}_{m,k}|^2$. With further manipulation, for a BER-target $\xi$, the maximum constellation size for user $n$ over subcarrier $k$ is obtained as 
\begin{align}
M^{*}_{n,k}(\Upsilon) = 1 + \frac{\zeta \Upsilon_{n,k} P_{n,k}(\Upsilon) }{\min \left(\frac{P_t}{K} , \frac{I^{m}_{th}}{N^{sp}_{m}} \right)} 
\label{optConstSize}
\end{align}
where 
\begin{align}
\zeta = \frac{-1.5}{\ln(\xi/0.3)}.
\end{align}

According the constraints (\ref{OF1b}) and (\ref{OF1c}) in the optimization problem $\mathscr{O}_{1}$, the cumulative density function (cdf) of $\gamma_{n,k}$ can be written
\begin{align}
F_{\gamma_{n,k}}(\Gamma) = \mathscr{P} \biggr( \frac{P_{t} |H^{ss}_{n,k}|^2}{K (\sigma^2_{n} + \sigma^2_{ps})} \leq \Gamma , \frac{I^{m}_{th} |H^{ss}_{n,k}|^2}{N^{sp}_{m} (\sigma^2_{n} + \sigma^2_{ps})} \leq \Gamma \biggr).
\label{cdfofG}
\end{align}
The probability expression in (\ref{cdfofG}) can be further simplified by considering the cases $\frac{P_{t} |H^{ss}_{n,k}|^2}{K (\sigma^2_{n} + \sigma^2_{ps})} \lesseqqgtr \frac{I^{m}_{th} |H^{ss}_{n,k}|^2}{N^{sp}_{m} (\sigma^2_{n} + \sigma^2_{ps})}$ and conditioning on $N^{sp}_{m}$
\begin{align}
& 1 - \mathscr{P} \biggr( \frac{P_{t} |H^{ss}_{n,k}|^2}{K (\sigma^2_{n} + \sigma^2_{ps})} > \Gamma , \frac{I^{m}_{th} |H^{ss}_{n,k}|^2}{N^{sp}_{m} (\sigma^2_{n} + \sigma^2_{ps})} > \Gamma \biggr) = \nonumber \\ & 1 -
  \begin{dcases}
   \mathscr{P} \biggr( |H^{ss}_{n,k}|^2 > \frac{K \Gamma (\sigma^2_{n} + \sigma^2_{ps})}{P_{t}} \biggr) & N^{sp}_{m} \leq \frac{I^{m}_{th} K}{P_t} \nonumber \\
   \mathscr{P} \biggr( |H^{ss}_{n,k}|^2 > \frac{N^{sp}_{m} \Gamma (\sigma^2_{n} + \sigma^2_{ps})}{I^{m}_{th}} \biggr) & N^{sp}_{m} > \frac{I^{m}_{th} K}{P_t}.
  \end{dcases}\\
  \label{sigmaOPT}
\end{align}

\textit{Lemma 1:}
For large values of $K$, given complex Gaussian random variables $H^{sp}_{m,k}$ with means $\mu_{H^{sp}_{m,k}}$ and equal variance $\delta^{2}_{H^{sp}_{m,k}}$ for all $k \in \{1, ..., K\}$, the non-central Chi-square random variable $N^{sp}_{m} = \sum_{k=1}^{K} |H^{sp}_{m,k}|^2$ can be approximated as a Gaussian random variable with respective mean and variance $\mu_{N^{sp}_{m}} \! = \! \delta_{H^{sp}_{m,k}}^{2} \left[ 2K \! + \! \mu^{'} \right]$ and\linebreak $\delta^{2}_{N^{sp}_{m}} \! = \! \delta^4_{H^{sp}_{m,k}} \left[ 4K \! + \! 4 \mu^{'} \right]$, where ${\mu^{'}} = \sum_{k=1}^{K}(\frac{\mu_{H^{sp}_{m,k}}}{\delta_{H^{sp}_{m,k}}})^2$.\\

\textit{Proof 1:} We can write $H^{sp}_{m,k} = \delta_{H^{sp}_{m,k}} G^{sp}_{m,k}$, where $G^{sp}_{m,k} \thicksim CN(\frac{\mu_{H^{sp}_{m,k}}}{\delta_{H^{sp}_{m,k}}},1)$. Assuming equal variance for random variables $H^{sp}_{m,k}$, $\sum_{k=1}^{K} |G^{sp}_{m,k}|^2$ is a non-central Chi-Square random variable with degree of freedom $2K$ and non-centrality parameter $\mu^{'} = \sum_{k=1}^{K}(\frac{\mu_{H^{sp}_{m,k}}}{\delta_{H^{sp}_{m,k}}})^2$.
For large values of $K$, central limit theorem (CLT) can be invoked to show that the non-central Chi-Square random variable $\sum_{k=1}^{K} |G^{sp}_{m,k}|^2$, can be approximated as a Gaussian random variable as follows
\begin{align}
\sum_{k=1}^{K} |G^{sp}_{m,k}|^{2} \thicksim N \left( 2 K \! + \! \mu^{'},4K \! + \! 4 \mu^{'} \right).
\label{Gsp1}
\end{align} 
Hence, $N^{sp}_{m} = \sum_{k=1}^{K} |H^{sp}_{m,k}|^2$ can be approximated by
\begin{align}
N^{sp}_{m} = \sum_{k=1}^{K} |H^{sp}_{m,k}|^2 \thicksim N \biggr( \mu_{N^{sp}_{m}} , \delta^{2}_{N^{sp}_{m}} \biggr)
\label{NCCtoN}
\end{align}
where $\mu_{N^{sp}_{m}} = \delta^2_{H^{sp}_{m,k}} \Big[ 2K + \mu^{'} \Big]$ and $\delta^{2}_{N^{sp}_{m}} = \delta^{4}_{H^{sp}_{m,k}} \Big[ 4K + 4 \mu^{'} \Big]$. 
Denoting the pdf of $N^{sp}_{m}$ with $f_{N^{sp}_{m}}(.)$, and the cdfs of $|H^{ss}_{n,k}|^2$ and $N^{sp}_{m}$ with $F_{|H^{ss}_{n,k}|^2}(.)$ and $F_{N^{sp}_{m}}(.)$, respectively, we write the cdf of $\gamma_{n,k}$ as
\begin{align}
F_{\gamma_{n,k}}(\Gamma) & = 1 -  A - B,
\end{align}
\begin{align}
& A = \!\! \int^{\frac{I^{m}_{th} K}{P_t}}_{0} \!\!\!\! \mathscr{P} \! \left(\! |H^{ss}_{n,k}|^2 \! > \! \frac{K \Gamma (\sigma^2_{n} + \sigma^2_{ps})}{P_{t}} \! \right) \! f_{N^{sp}_{m}}(N^{sp}_{m}) d N^{sp}_{m} \nonumber \\ &  = \! \mathscr{P} \! \left(\! |H^{ss}_{n,k}|^2 \! > \! \frac{K \Gamma (\sigma^2_{n} + \sigma^2_{ps})}{P_{t}} \! \right) \int^{\frac{I^{m}_{th} K}{P_t}}_{0} \!\!\!\! f_{N^{sp}_{m}}(N^{sp}_{m}) \, d N^{sp}_{m} \nonumber \\ & = \mathscr{P} \! \left(\! |H^{ss}_{n,k}|^2 \! > \! \frac{K \Gamma (\sigma^2_{n} + \sigma^2_{ps})}{P_{t}} \! \right) \mathscr{P} \left( N^{sp}_{m} \leq \frac{I^{m}_{th} K}{P_t} \right) \nonumber \\ & = \left( 1 - F_{|H_{ss}|^2} \biggr( \frac{K \Gamma (\sigma^2_{n} + \sigma^2_{ps})}{P_{t}} \biggr) \right) \, F_{N^{sp}_{m}} \left( \frac{I^{m}_{th} K}{P_t} \right) 
\label{integA}
\end{align}
and
\begin{align}
\! B \! = \!\!\! \int_{\frac{I^{m}_{th} K}{P_t}}^{\infty} \!\! \mathscr{P} \! \biggr(\! |H^{ss}_{n,k}|^2 \!\! > \!\! \frac{N^{sp}_{m} \Gamma (\sigma^{2}_{n} \! + \! \sigma^{2}_{ps})}{I^{m}_{th}} \! \biggr) f_{N^{sp}_{m}}(N^{sp}_{m}) \, d N^{sp}_{m}.
\label{integB}
\end{align}
Recall that the cdf of a Normally-distributed random variable $X$ with mean $\mu$ and standard deviation $\sigma$ is given by $F_{X}(x)=\frac{1}{2} \left[ 1 + erf \left( \frac{x - \mu}{2 \sigma^2} \right) \right]$, and the cdf of an Exponentially-distributed random variable $Y$ is computed by $F_{Y}(y) = 1 - e^{-y/\mu}$, where $\mu$ is the mean. Suppose that $|H^{ss}_{n,k}|^2$ follows an exponential distribution with mean $\mu_{|H^{ss}_{n,k}|^2}$, hence, the integrals in (\ref{integA}) and (\ref{integB}) can be simplified to (\ref{simpleA}) and (\ref{simpleB}), respectively. 
\begin{figure*}[!t]
\normalsize
\setcounter{equation}{18}
\begin{align}
A & \! = \! \frac{1}{2} \exp \left( \frac{- K \Gamma (\sigma^2_{n} + \sigma^2_{ps})}{P_t \mu_{|H^{ss}_{n,k}|^2}} \right) \Biggr[1 + erf \biggr( \frac{\frac{I^{m}_{th} K}{P_t} - \mu_{N^{sp}_{m}}}{\sqrt{2 \delta^{2}_{N^{sp}_{m}}}} \biggr) \Biggr]
\label{simpleA}
\end{align}
\begin{align}
& B \! = \! \int_{\frac{I^{m}_{th} K}{P_t}}^{\infty} \!\! \frac{\exp \left(\! \frac{- N^{sp}_{m} \Gamma (\sigma^2_{n} + \sigma^2_{ps})}{\mu_{|H^{ss}_{n,k}|^2} I^{m}_{th}} \! \right) \exp \left(\! \frac{- (N^{sp}_{m} - \mu_{N^{sp}_{m}})^2}{2 \delta^{2}_{N^{sp}_{m}}} \! \right) }{\sqrt{2 \pi \delta^{2}_{N^{sp}_{m}}}} d N^{sp}_{m} \approx \frac{1}{2} \exp \Biggr( {\frac{\Gamma (\sigma^2_{n} + \sigma^2_{ps}) (-2 \mu_{N^{sp}_{m}} \mu_{|H^{ss}_{n,k}|^2} I^{m}_{th} + \delta^{2}_{N^{sp}_{m}} \Gamma (\sigma^2_{n} + \sigma^2_{ps}))}{2 \mu^2_{H^{ss}_{n,k}} {I^{m}_{th}}^2 }} \Biggr) \nonumber \\ & \Biggr[ 1 - erf \biggr( \frac{ \mu_{|H^{ss}_{n,k}|^2} I^{m}_{th} \left( - \mu_{N^{sp}_{m}} + \frac{I^{m}_{th} K}{P_{t}} \right) + \delta^{2}_{N^{sp}_{m}} \Gamma (\sigma^2_{n} + \sigma^2_{ps})}{\sqrt{2} \mu_{|H^{ss}_{n,k}|^2} I^{m}_{th} \delta_{N^{sp}_{m}}} \biggr) \Biggr].
\label{simpleB} 
\end{align}
\begin{align}
& F_{\gamma_{n,k}}(\Gamma) \approx \nonumber \\ & 1 - \frac{1}{2} \exp \left(\! \frac{- K \Gamma (\sigma^2_{n} + \sigma^2_{ps})}{P_t \mu_{|H^{ss}_{n,k}|^2}} \! \right) \Biggr[1 + erf \biggr( \frac{\frac{I^{m}_{th} K}{P_t} - \mu_{N^{sp}_{m}}}{\sqrt{2 \delta^{2}_{N^{sp}_{m}}}} \biggr) \Biggr] - \frac{1}{2} \exp \Biggr(\! {\frac{\Gamma (\sigma^2_{n} + \sigma^2_{ps}) (-2 \mu_{N^{sp}_{m}} \mu_{|H^{ss}_{n,k}|^2} I^{m}_{th} + \delta^{2}_{N^{sp}_{m}} \Gamma (\sigma^2_{n} + \sigma^2_{ps}))}{2 \mu^2_{H^{ss}_{n,k}} {I^{m}_{th}}^2}} \! \Biggr) \nonumber \\ & \Biggr[ 1 - erf \biggr( \frac{ \mu_{|H^{ss}_{n,k}|^2} I^{m}_{th} \left( - \mu_{N^{sp}_{m}} + \frac{I^{m}_{th} K}{P_{t}} \right) + \delta^{2}_{N^{sp}_{m}} \Gamma (\sigma^2_{n} + \sigma^2_{ps})}{\sqrt{2} \mu_{|H^{ss}_{n,k}|^2} I^{m}_{th} \delta_{N^{sp}_{m}}} \biggr) \Biggr].
\label{CDFofgamma}    
\end{align}
\begin{align}
& f_{\gamma_{n,k}}(\Gamma) \approx \frac{K (\sigma^2_{n} + \sigma^2_{ps}) \exp \left( -\frac{K \Gamma (\sigma^2_{n} + \sigma^2_{ps}) }{P_t \mu_{|H^{ss}_{n,k}|^2}} \right) \left(erf \left(\frac{\frac{I^{m}_{th} K}{P_t}-\mu_{N^{sp}_{m}}}{\sqrt{2 \delta^{2}_{N^{sp}_{m}}}}\right)+1\right)}{2 P_t \mu_{|H^{ss}_{n,k}|^2} } \nonumber \\ & +\frac{(\sigma^2_{n} + \sigma^2_{ps}) \delta_{N^{sp}_{m}} \exp \left( -\frac{{I^{m}_{th}}^2 K^2 \mu_{|H^{ss}_{n,k}|^2}-2 I^{m}_{th} K \mu_{N^{sp}_{m}} \mu_{|H^{ss}_{n,k}|^2} P_t+2 K (\sigma^2_{n} + \sigma^2_{ps}) P_t \delta_{N^{sp}_{m}} \Gamma +{\mu^2_{N^{sp}_{m}}} \mu_{|H^{ss}_{n,k}|^2} P^2_t}{2 \mu_{|H^{ss}_{n,k}|^2} P^2_t \delta_{N^{sp}_{m}}} \right) }{\sqrt{2 \pi} I^{m}_{th} \mu_{|H^{ss}_{n,k}|^2}} \nonumber \\ & -\frac{\begin{aligned} & 0.5 (\sigma^2_{n} + \sigma^2_{ps}) (I^{m}_{th} \mu_{N^{sp}_{m}} \mu_{|H^{ss}_{n,k}|^2}-(\sigma^2_{n} + \sigma^2_{ps}) \delta_{N^{sp}_{m}} \Gamma) \exp \left( \frac{(\sigma^2_{n} + \sigma^2_{ps}) \Gamma ((\sigma^2_{n} + \sigma^2_{ps}) \delta_{N^{sp}_{m}} \Gamma-2 I^{m}_{th} \mu_{N^{sp}_{m}} \mu_{|H^{ss}_{n,k}|^2})}{2 {I^{m}_{th}}^2 {\mu_{|H^{ss}_{n,k}|^2}}^2} \right) \\ & \times \left(erf \left(\frac{ \left(I^{m}_{th} \mu_{|H^{ss}_{n,k}|^2} \left(\frac{I^{m}_{th} K}{P_t}-\mu_{N^{sp}_{m}}\right)+(\sigma^2_{n} + \sigma^2_{ps}) \delta_{N^{sp}_{m}} \Gamma \right)}{\sqrt{2 \delta^{2}_{N^{sp}_{m}}} I^{m}_{th} \mu_{|H^{ss}_{n,k}|^2} }\right)-1\right) \end{aligned}}{{I^{m}_{th}}^2 {\mu^{2}_{|H^{ss}_{n,k}|^2}}}
\label{PDFofgamma} 
\end{align}
\hrulefill
\vspace*{4pt}
\end{figure*}
Finally, a closed-form expression for cdf of $\gamma_{n,k}$ is developed in (\ref{CDFofgamma}). Trivially, through respective differentiation of (\ref{CDFofgamma}), the pdf of $\gamma_{n,k}$ is obtained in (\ref{PDFofgamma}). 

\subsection{Obtaining Solutions}

It can be observed that the optimization problem, $\mathscr{O}_{1}$, is convex with respect to the transmit power $P_{n,k}(\Upsilon)$, however, it is non-convex with respect to $\varphi_{n,k}(\Upsilon)$ as the time-sharing factor only takes binary values. To obtain a sub-optimal solution for problem $\mathscr{O}_{1}$, we employ the Lagrangian dual decomposition algorithm. By applying dual decomposition, the non-convex optimization problem, $\mathscr{O}_{1}$, is decomposed into independent sub-problems each corresponding to a given cognitive user.   

The Lagrangian function of problem $\mathscr{O}_{1}$ is expressed as\footnote{For simplicity, the analysis is carried out for a single primary receiver.}
\allowdisplaybreaks{
\begin{align}
& L(\varphi_{n,k}(\Upsilon),P_{n,k}(\Upsilon),\lambda(\Upsilon),\mu,\eta(\Upsilon)) = \nonumber \\ & \sum_{k=1}^{K} \sum_{n=1}^{N} E_{\Upsilon} \Bigg\{ \log_{2} \left( 1 + \frac{\zeta \Upsilon_{n,k} P_{n,k}(\Upsilon)}{\min \left(\frac{P_t}{K} , \frac{I^{m}_{th}}{N^{sp}_{m}} \right)} \right) \varphi_{n,k}(\Upsilon) \Bigg\} \nonumber \\ & -  \sum_{k=1}^{K} \lambda_{k}(\Upsilon) \Biggr( \sum_{n=1}^{N} \varphi_{n,k}(\Upsilon) - 1 \Biggr)  \nonumber \\ & - \mu \Biggr( E_{\Upsilon} \Big\{ \varphi_{n,k}(\Upsilon)  P_{n,k}(\Upsilon) \Big\} - P_t \Biggr) \nonumber \\ & - \eta(\Upsilon) \Biggr( \sum_{k=1}^{K} \sum_{n=1}^{N} \varphi_{n,k}(\Upsilon) P_{n,k}(\Upsilon) |H^{sp}_{m,k}|^2 - I^{m}_{th} \Biggr)  
\label{LagrangianFunction}
\end{align}
where $\mu$, $\eta(\Upsilon)$, and $\lambda_k(\Upsilon)$ are the non-negative Lagrangian multipliers. Define
\begin{align}
& l(\varphi_{n,k}(\Upsilon),P_{n,k}(\Upsilon),\lambda(\Upsilon),\mu,\eta(\Upsilon)) = \nonumber \\ & \sum_{k=1}^{K} \sum_{n=1}^{N} \log_{2} \left( 1 + \frac{\zeta \Upsilon_{n,k} P_{n,k}(\Upsilon)}{\min \left(\frac{P_t}{K} , \frac{I^{m}_{th}}{N^{sp}_{m}} \right)} \right) \varphi_{n,k}(\Upsilon) \nonumber \\ &  -  \sum_{k=1}^{K} \lambda_{k}(\Upsilon) \Biggr( \sum_{n=1}^{N} \varphi_{n,k}(\Upsilon) - 1 \Biggr)  \nonumber \\ & - \mu \Biggr(  \varphi_{n,k}(\Upsilon)  P_{n,k}(\Upsilon) - P_t \Biggr) \nonumber \\ & - \eta(\Upsilon) \Biggr( \sum_{k=1}^{K} \sum_{n=1}^{N} \varphi_{n,k}(\Upsilon) P_{n,k}(\Upsilon) |H^{sp}_{m,k}|^2 - I^{m}_{th} \Biggr).  
\label{SmallL}
\end{align}
Note that the variation of the Lagrangian function, $L(\varphi_{n,k}(\Upsilon),P_{n,k}(\Upsilon),\lambda(\Upsilon),\mu,\eta(\Upsilon))$, in (\ref{LagrangianFunction}) with respect to the optimization parameters, $\varphi_{n,k}(\Upsilon)$ and $P_{n,k}(\Upsilon)$, is equal to zero if and only if the derivative of $l(\varphi_{n,k}(\Upsilon),P_{n,k}(\Upsilon),\lambda(\Upsilon),\mu,\eta(\Upsilon))$ with respect to $\varphi_{n,k}(\Upsilon)$ and $P_{n,k}(\Upsilon)$ is equal to zero \cite{1459060}. 

Based on the Karush-Kuhn-Tucker (KKT) necessary conditions theorem, the optimum solutions $\left( P^{*}_{n,k}(\Upsilon), \varphi^{*}_{n,k}(\Upsilon) \right)$ must satisfy the following
conditions:
\begin{align}
\frac{\partial l(\varphi_{n,k}(\Upsilon),P_{n,k}(\Upsilon),\lambda(\Upsilon),\mu,\eta(\Upsilon))}{\partial P_{n,k}(\Upsilon)} 
\begin{dcases}
     = 0, \; P_{n,k}(\Upsilon) > 0 \\
      < 0, \; P_{n,k}(\Upsilon) = 0
\end{dcases}\\
\frac{\partial l(\varphi_{n,k}(\Upsilon),P_{n,k}(\Upsilon),\lambda(\Upsilon),\mu,\eta(\Upsilon))}{\partial \varphi_{n,k}(\Upsilon)} 
\begin{dcases}
     < 0, \; \varphi_{n,k}(\Upsilon) = 0 \\
     = 0, \; \varphi_{n,k}(\Upsilon) \! \in \! (0,1)\\
     > 0, \; \varphi_{n,k}(\Upsilon) = 1
\end{dcases}
\end{align}
\vspace*{-1em}
\begin{align}
\lambda_{k}(\Upsilon) \Biggr( \sum_{n=1}^{N} \varphi_{n,k}(\Upsilon) - 1 \Biggr) = 0 \\
\mu \Biggr( E_{\Upsilon} \Big\{ \varphi_{n,k}(\Upsilon)  P_{n,k}(\Upsilon) \Big\} - P_t \Biggr) = 0\\
\eta(\Upsilon) \Biggr( \sum_{k=1}^{K} \sum_{n=1}^{N} \varphi_{n,k}(\Upsilon) P_{n,k}(\Upsilon) |H^{sp}_{m,k}|^2 - I^{m}_{th} \Biggr) = 0
\end{align}
The Lagrangian dual optimization problem associated with (\ref{LagrangianFunction}) is given by}
\begin{align}
\min_{\lambda(\Upsilon),\mu,\eta(\Upsilon)}(F \left( \lambda(\Upsilon),\mu,\eta(\Upsilon) \right) \, , \, \text{s.t.:} \, \lambda(\Upsilon),\mu,\eta(\Upsilon) \geq 0
\label{nonDiffDual}
\end{align}
where $F(\lambda(\Upsilon),\mu,\eta(\Upsilon))$ denotes the Lagrangian dual function formulated below
\begin{align}
F(\lambda(\Upsilon),\mu,\eta(\Upsilon)) & = \sum_{n=1}^{N} f_{n}(\varphi_{n,k}(\Upsilon),P_{n,k}(\Upsilon)) + \sum_{k=1}^{K} \lambda_{k}(\Upsilon) \nonumber \\ & + \mu P_{t} + \eta(\Upsilon) I^{m}_{th}
\label{dualFUNCOPT1}
\end{align}
where
\begin{align}
& f_{n}(\varphi_{n,k}(\Upsilon),P_{n,k}(\Upsilon)) = \nonumber \\ & \max_{\varphi_{n,k}(\Upsilon),P_{n,k}(\Upsilon)} \Biggr( \sum_{k=1}^{K} \log_{2} \Biggr( 1 + \frac{\zeta \Upsilon_{n,k} P_{n,k}(\Upsilon)}{\min \left(\frac{P_t}{K} , \frac{I^{m}_{th}}{N^{sp}_{m}} \right)} \Bigg) \varphi_{n,k}(\Upsilon)  \nonumber \\ & - \sum_{k=1}^{K} \lambda_{k}(\Upsilon) \varphi_{n,k}(\Upsilon) - \mu \sum_{k=1}^{K}  \varphi_{n,k}(\Upsilon) P_{n,k}(\Upsilon) \! \nonumber \\ & - \eta(\Upsilon) \sum_{k=1}^{K} \varphi_{n,k}(\Upsilon) P_{n,k}(\Upsilon) |H^{sp}_{m,k}|^2 \Biggr).
\label{LFNSFORN1}
\end{align}
To find the optimum solution of problem (\ref{LFNSFORN1}), we differentiate $f_{n}(\varphi_{n,k}(\Upsilon),P_{n,k}(\Upsilon))$ with respect to $\varphi_{n,k}(\Upsilon) P_{n,k}(\Upsilon)$
\begin{align}
\frac{\partial f_{n}(\varphi_{n,k}(\Upsilon),P_{n,k}(\Upsilon))}{\partial (\varphi_{n,k}(\Upsilon) P_{n,k}(\Upsilon))} & = \frac{\frac{\zeta \Upsilon_{n,k}}{\min \left(\frac{P_t}{K} , \frac{I^{m}_{th}}{N^{sp}_{m}} \right)}}{\ln(2) \left( 1 + \frac{\zeta \Upsilon_{n,k} P_{n,k}(\Upsilon) }{\min \left(\frac{P_t}{K} , \frac{I^{m}_{th}}{N^{sp}_{m}} \right)} \right)} \nonumber \\ & - \mu - \eta(\Upsilon) |H^{sp}_{m,k}|^2.  
\end{align}
Applying the KKT conditions yields the optimal potential power allocation policy for Lagrangian multipliers $\mu$ and $\eta(\Upsilon)$
\begin{align}
P^{*}_{n,k}(\Upsilon) = \Biggr[\frac{1}{\ln(2) (\mu +  \eta(\Upsilon) |H^{sp}_{m,k}|^2)} - \frac{\min \left(\frac{P_t}{K} , \frac{I^{m}_{th}}{N^{sp}_{m}} \right)}{\zeta \Upsilon_{n,k}} \Biggr]^{+} 
\label{Pstar}
\end{align}
where $[x]^{+} \triangleq \max\{x,0\}$. The solution in (\ref{Pstar}) can be considered as a multi-level water-filling algorithm where each subcarrier has a distinct water-level for a given user. Note that the water levels determine the potential optimum amount of power that may be allocated to $n^\text{th}$ CRx over subcarrier $k$.
The result in (\ref{Pstar}) can be used to find the optimal subcarrier allocation strategy. By differentiating $f_{n}(\varphi_{n,k}(\Upsilon),P_{n,k}(\Upsilon))$ with respect to $\varphi_{n,k}(\Upsilon)$ we have
\begin{align}
& \frac{\partial f_{n}(\varphi_{n,k}(\Upsilon),P_{n,k}(\Upsilon))}{\partial \varphi_{n,k}(\Upsilon)} = \frac{\frac{\zeta \Upsilon_{n,k} P^{*}_{n,k}(\Upsilon)}{\min \left(\frac{P_t}{K} , \frac{I^{m}_{th}}{N^{sp}_{m}} \right)}}{\ln(2) \left( 1 + \frac{\zeta \Upsilon_{n,k} P^{*}_{n,k}(\Upsilon)}{\min \left(\frac{P_t}{K} , \frac{I^{m}_{th}}{N^{sp}_{m}} \right)} \right)} \nonumber \\ & + \frac{\ln \left( 1 + \frac{\zeta \Upsilon_{n,k} P^{*}_{n,k}(\Upsilon)}{\min \left(\frac{P_t}{K} , \frac{I^{m}_{th}}{N^{sp}_{m}} \right)} \right)}{\ln(2)} - \lambda_{k}(\Upsilon). 
\label{diffVARPHI1} 
\end{align}
By substituting the optimal power policy (\ref{Pstar}) in (\ref{diffVARPHI1}) and by applying the KKT conditions, the optimal subcarrier allocation problem is 
formulated as:
\begin{align}
n^{*} = argmax(\Lambda(\Upsilon_{n,k})) \; , \; \forall n \in \{1,...,N\} \; , \; \forall k \in \{1,...,K\}\nonumber \\
\label{PROBLEMSCAPOPT}
\end{align}
where $n^{*}$ is the optimal CRx index, and
\begin{align}
& \Lambda(\Upsilon_{n,k}) = \nonumber \\ & \frac{\frac{\zeta \Upsilon{n,k} P^{*}_{n,k}(\Upsilon) }{\min \left(\frac{P_t}{K} , \frac{I^{m}_{th}}{N^{sp}_{m}} \right)}}{\ln(2) \left( 1 + \frac{\zeta \Upsilon{n,k} P^{*}_{n,k}(\Upsilon)}{\min \left(\frac{P_t}{K} , \frac{I^{m}_{th}}{N^{sp}_{m}} \right)} \right)} + \frac{\ln \left( 1 + \frac{\zeta \Upsilon{n,k} P^{*}_{n,k}(\Upsilon)}{\min \left(\frac{P_t}{K} , \frac{I^{m}_{th}}{N^{sp}_{m}} \right)} \right)}{\ln(2)}.
\label{SCAPOPT}
\end{align}
The optimal subcarrier allocation policy is therefore achieved by assigning the $k^{th}$ subcarrier to the user with the highest value of $\Lambda(\Upsilon_{n,k})$ for all corresponding $\gamma_{n,k}$. To ensure optimality, $\lambda_{k}(\Upsilon)$ should be between first and second maximas of $\Lambda(\Upsilon_{n,k})$. If there are multiple equal maximas, the time-slot can be identically shared among the respective users. Substituting (\ref{Pstar}) and (\ref{SCAPOPT}) in (\ref{LFNSFORN1}), derives $f_{n}(\varphi_{n,k}(\Upsilon),P_{n,k}(\Upsilon))$, therefore, the solution for (\ref{dualFUNCOPT1}) can be obtained. To compute the solution for the non-differentiable dual problem in    (\ref{nonDiffDual}), different optimization algorithms can be applied, including subgradient, ellipsoid, and cutting-plane. \linebreak In this work, we use the subgradient-based method to update the values of the coefficients $\lambda_{k}(\Upsilon)$, $\mu$, and $\eta(\Upsilon)$, in order to determine the optimal solution to (\ref{nonDiffDual}).

The subgradient method has been widely used for solving Lagrangian relaxation problems. The master problem sets the user resource allocation prices, and in order to update the dual variables, in every iteration of the subgradient method, the algorithm repeatedly finds the maximizing assignment for the sub-problems individually. For any optimal pair of $(\varphi^{*}_{n,k}(\Upsilon),P^{*}_{n,k}(\Upsilon))$, the dual variables of problem (\ref{dualFUNCOPT1}) are updated using the following iterations
\begin{gather}
\mu^{i+1} = \mu^{i} - \tau^{i}_{1} \left( P_{t} -  \sum_{n=1}^{N} \sum_{k=1}^{K} E_{\Upsilon} \left\{ \varphi^{*}_{n,k}(\Upsilon) P^{*}_{n,k}(\Upsilon) \right\} \right) 
\label{SUB1} \\
\eta^{i+1}(\Upsilon) = \eta^{i}(\Upsilon) - \tau^{i}_{2} \nonumber \\ \times \left( I^{m}_{th} - \sum_{n=1}^{N} \sum_{k=1}^{K} \varphi^{*}_{n,k}(\Upsilon) P^{*}_{n,k}(\Upsilon) |H^{sp}_{m,k}|^2 \right) \label{SUB2}
\end{gather}
where for the iteration number $i$, $\tau^{i}_{1}$ and $\tau^{i}_{2}$ are the step sizes. The initial values of dual multipliers and step size selection are important towards obtaining the optimal solution, and can greatly effect the optimization problem convergence. 

\begin{algorithm}[t]
\caption{Subgradient-based method; $ASE^{*}$, $M^{*}_{n,k}(\Upsilon)$, $\varphi^{*}_{n,k}(\Upsilon)$, and $P^{*}_{n,k}(\Upsilon)$, are the optimal values of $ASE$, $M_{n,k}(\Upsilon)$, $\varphi_{n,k}(\Upsilon)$, and $P_{n,k}(\Upsilon)$, respectively.
\label{alg:Subgradiant-method}}
\begin{enumerate}
\item Assign initial values to $\lambda_{k}(\Upsilon)$, $\mu$, $\eta(\Upsilon)$, $\tau^{i}_{1}$, and $\tau^{i}_{2}$, $\forall n \in \{1, ..., N\}$, and $\forall k \in \{1, ..., K\}$, respectively.
\item Calculate $P_{n,k}(\Upsilon)$ and $\varphi_{n,k}(\Upsilon)$, $\forall n \in \{1, ..., N\}$, and $\forall k \in \{1, ..., K\}$, using (\ref{Pstar}) and (\ref{PROBLEMSCAPOPT}), respectively.
\item Update $\lambda_{k}(\Upsilon)$, $\mu$, $\eta(\Upsilon)$, $\tau^{i}_{1}$, and $\tau^{i}_{2}$, for any $n \in \{1, ..., N\}$, and $k \in \{1, ..., K\}$, according to (\ref{SUB1}) and (\ref{SUB2}). 
\item Repeat steps 2 and 3 until convergence.
\item Determine $P^{*}_{n,k}(\Upsilon)$ and $\varphi^{*}_{n,k}(\Upsilon)$, using (\ref{Pstar}) and (\ref{PROBLEMSCAPOPT}), respectively.
\item Based on the obtained result from step 5, calculate  $M^{*}_{n,k}(\Upsilon)$ using (\ref{optConstSize}), hence, compute $ASE^{*}$ according to (\ref{AASE}).
\end{enumerate}
\end{algorithm}

The potential optimum continuous-rate adaptive constellation size vector for user $n$ over subcarrier $k$ is written as
\begin{align}
& M^{*}_{n,k}(\Upsilon) = \nonumber \\ & \max \Biggl(1,\frac{\zeta \Upsilon_{n,k}}{\ln(2) \min \left(\frac{P_t}{K} , \frac{I^{m}_{th}}{N^{sp}_{m}} \right) (\mu + \eta(\Upsilon) |H^{sp}_{m,k}|^2} \Biggl).
\label{optimumConstellation}
\end{align}
Note that the aforementioned expression serves as an upper-bound for practical scenarios where only discrete-valued constellation sizes are applicable. Nevertheless, the real-valued $M^{*}_{n,k}(\Upsilon)$ in (\ref{optimumConstellation}) may be truncated to the nearest integer. The corresponding maximum aggregate average spectral efficiency of the adaptive MQAM/OFDMA system is thus derived below
\begin{align}
& ASE^{*} = \sum_{n=1}^{N} \sum_{k=1}^{K} E_{\Upsilon} \Bigg\{   \log_{2} \Biggl[ \max \Biggl( 1, \nonumber \\ &  \frac{\zeta \Upsilon_{n,k}}{\ln(2) \min \left(\frac{P_t}{K} , \frac{I^{m}_{th}}{N^{sp}_{m}} \right) (\mu + \eta(\Upsilon) |H^{sp}_{m,k}|^2)} \Biggl) \Biggl] \varphi^{*}_{n,k}(\Upsilon) \Bigg\}.
\end{align}
According to (\ref{optimumConstellation}), no transmission takes place, i.e., $M^{*}_{n,k}(\Upsilon)=1$, when $P^{*}_{n,k}(\Upsilon) = 0$. Consequently, the optimized cut-off threshold, dictated by the channel quality, power constraint, and interference constraint, is given by: $\Upsilon^{th}_{n,k} = \frac{\ln(2) (\mu + \eta(\Upsilon) |H^{m}_{sp}|^2) \min \left(\frac{P_t}{K} , \frac{I^{m}_{th}}{N^{sp}_{m}} \right)}{\zeta}$.

\section{Interference Constraint with Average Case/Worst Case Imperfect Cross-Link CSI}

Due to technical reasons such as estimation errors and wireless channel delay, perfect channel information is not available. In shared-spectrum environments, controlling the interference on the primary receivers is highly dependent on the accuracy of the cross-service channel estimation. Here, we assume that imperfect cross-link knowledge between CTx and PRxs is available at the secondary transmitter. The interference management at the cognitive base station is based on the noisy estimation of CTx and PRx channel-to-noise-plus-interference ratio (CINR) by $\hat{H}^{sp}_{m,k}$ in (\ref{errorF}). As previously mentioned, by considering the `correlated case' of the estimation $\hat{H}^{sp}_{m,k}$ and error $\Delta H^{sp}_{m,k}$ random variables, we derive a \textit{posterior} distribution of the actual channel conditioned on its estimate, to facilitate robust and reliable interference management.

The maximum achievable aggregate spectral efficiency in bits per second per Hertz (bps/Hz), for the cognitive radio system operating under peak aggregate interference constraint and total average transmit power constraint, for a given BER-target quality, with noisy cross-link CSI, is the solution to the following optimization problem.

\textit{Problem} $\mathscr{O}_{2}$:
\begin{subequations}
\label{optimizationProb2}
\begin{gather}
\max_{\varphi_{n,k}(\Upsilon),P_{n,k}(\Upsilon)} \sum_{n=1}^{N} \sum_{k=1}^{K} E_{\Upsilon | \hat{h}^{sp}} \Big\{ \log_{2}(M_{n,k}(\Upsilon)) \varphi_{n,k}(\Upsilon) \Big\} \label{OF2a}\\ 
\text{s. t.:} \quad \text{constraints in (\ref{OF1b}), (\ref{OF1d}), (\ref{OF1be}), and (\ref{OF1e})}, \nonumber \\
\sum_{n=1}^{N} \sum_{k=1}^{K} \varphi_{n,k}(\Upsilon) P_{n,k}(\Upsilon) |H^{sp}_{m,k} | \hat{H}^{sp}_{m,k}|^2 \! \leq \! I^{m}_{th} , \forall m \! \in \! \{1, ..., M\} \label{OF2c}
\end{gather}
\end{subequations}
where $\hat{h}^{sp}$ is defined as a vector containing $\hat{H}^{sp}_{m,k}$ of all time intervals. The objective of this section is to devise an estimation framework by employing a posteriori pdf of the channel estimation error given the channel estimation. This general framework enables us to formulate the `average case' and `worst case' scenarios of the channel estimation error.  

\subsection{Analysis for the Average Case of Estimation Error}

\textit{Proposition 1:}
Given $\hat{H}^{sp}_{m,k}$ and $\Delta H^{sp}_{m,k}$ are statistically correlated random variables with a correlation factor $\rho = \sqrt{\delta^{2}_{\Delta H^{sp}_{m,k}}/(\delta^{2}_{\Delta H^{sp}_{m,k}} + \delta^{2}_{H^{sp}_{m,k}})}$, where $0 \leq \rho \leq 1$, hence, $cov(\hat{H}^{sp}_{m,k},\Delta H^{sp}_{m,k}) = \delta^{2}_{\Delta H^{sp}_{m,k}}$\footnote{$var(x)$ denotes the variance of $x$ and $cov(y,z)$ is defined as the covariance of $y$ and $z$.}. The posterior distribution of $\Delta H^{sp}_{m,k}$ given $\hat{H}^{sp}_{m,k}$ is a complex Gaussian random variable with respective mean and variance of
\begin{align}
& \mu_{\Delta H^{sp}_{m,k} | \hat{H}^{sp}_{m,k}} = E_{\Delta H^{sp}_{m,k} | \hat{H}^{sp}_{m,k}}(\Delta H^{sp}_{m,k} | \hat{H}^{sp}_{m,k}) \nonumber \\ & = E_{\Delta H^{sp}_{m,k}}(\Delta H^{sp}_{m,k}) + \frac{cov(\Delta H^{sp}_{m,k},\hat{H}^{sp}_{m,k})}{\delta^{2}_{\Delta H^{sp}_{m,k}} + \delta^{2}_{H^{sp}_{m,k}}} \nonumber \\ & \times \left( \hat{H}^{sp}_{m,k} - E_{\hat{H}^{sp}_{m,k}}(\hat{H}^{sp}_{m,k}) \right) = \rho^{2} \hat{H}^{sp}_{m,k} 
\label{ExpDaltaHhat}
\end{align}
and
\begin{align}
\delta^2_{\Delta H^{sp}_{m,k} | \hat{H}^{sp}_{m,k}} & = var(\Delta H^{sp}_{m,k} | \hat{H}^{sp}_{m,k}) \nonumber \\ &  = \delta^{2}_{\Delta H^{sp}_{m,k}} \biggr[1 - \frac{cov^{2}(\Delta H^{sp}_{m,k},\hat{H}^{sp}_{m,k})}{\delta^{2}_{\Delta H^{sp}_{m,k}} \, \delta^{2}_{\hat{H}^{sp}_{m,k}}} \biggr] \nonumber \\ &  = (1 - \rho^{2}) \delta^{2}_{\Delta H^{sp}_{m,k}}.
\end{align}

Using (\ref{errorF}), the interference constraint in (\ref{OF2c}) for the `average case' of estimation error can be written as
\begin{gather}
\sum_{n=1}^{N} \sum_{k=1}^{K} \varphi_{n,k}(\Upsilon) P_{n,k}(\Upsilon) \nonumber \\ \times \left( |\hat{H}^{sp}_{m,k} + \Delta H^{sp}_{m,k} | \hat{H}^{sp}_{m,k}| \right)^2 \leq I^{m}_{th}.
\end{gather}
With further analysis, the above is reduced to
\begin{gather}
\sum_{n=1}^{N} \sum_{k=1}^{K} \varphi_{n,k}(\Upsilon) P_{n,k}(\Upsilon) \nonumber \\ \left( |\hat{H}^{sp}_{m,k} | \hat{H}^{sp}_{m,k} + \Delta H^{sp}_{m,k} | \hat{H}^{sp}_{m,k} | \right)^2 \leq I^{m}_{th}
\end{gather}
where $\hat{H}^{sp}_{m,k} | \hat{H}^{sp}_{m,k}$ is a constant. Thus, by substituting the expectation in (\ref{ExpDaltaHhat}), we have
\begin{align}
\sum_{n=1}^{N} \sum_{k=1}^{K} \varphi_{n,k}(\Upsilon) P_{n,k}(\Upsilon) \left( | \hat{H}^{sp}_{m,k} + \rho^2 \hat{H}^{sp}_{m,k} | \right)^2 \leq I^{m}_{th}.
\end{align}

By adopting a similar approach to that in the previous section, we employ the Lagrangian dual optimization method to obtain $ASE^{*}$ for the `average case' scenario. The potential optimum power allocation policy for user $n$ and subcarrier $k$ is given by
\begin{align}
& P^{*}_{n,k}(\Upsilon) = \nonumber \\ & \Biggr[\frac{1}{\ln(2) (\mu + \eta(\Upsilon) |\hat{H}^{sp}_{m,k} (1 + \rho^2)|^2)} - \frac{{\min \left(\frac{P_t}{K} , \frac{I^{m}_{th}}{N^{sp}_{m}} \right)}}{\zeta \Upsilon_{n,k}} \Biggr]^{+}
\label{Pstar2}
\end{align}
where in the `average case', $N^{sp}_{m} = \sum_{k=1}^{K} ( | \hat{H}^{sp}_{m,k} ( 1 + \rho^2 | )^2$. 
The optimal subcarrier allocation policy is the solution to the following problem
\begin{align}
n^{*} = argmax(\Lambda(\Upsilon_{n,k})) \; , \; \forall n \in \{1,...,N\} \; , \; \forall k \in \{1,...,K\}
\label{PROBLEMSCAPOPT2}
\end{align}
where $n^{*}$ is the optimal CRx index, and
\begin{align}
& \Lambda(\Upsilon_{n,k}) = \nonumber \\ & \frac{\frac{\zeta \Upsilon_{n,k} P^{*}_{n,k}(\Upsilon)}{{\min \left(\frac{P_t}{K} , \frac{I^{m}_{th}}{N^{sp}_{m}} \right)}}}{\ln(2) \left( 1 + \frac{\zeta \Upsilon_{n,k} P^{*}_{n,k}(\Upsilon)}{{\min \left(\frac{P_t}{K} , \frac{I^{m}_{th}}{N^{sp}_{m}} \right)}} \right)} + \frac{\ln \left( 1 + \frac{\zeta \Upsilon_{n,k} P^{*}_{n,k}(\Upsilon)}{{\min \left(\frac{P_t}{K} , \frac{I^{m}_{th}}{N^{sp}_{m}} \right)}} \right)}{\ln(2)}.
\label{SCAPOPT2}
\end{align}
Subsequently, the optimal continuous-rate solution for the constellation size of user $n$ over subcarrier $k$ is derived
\begin{align}
& M^{*}_{n,k}(\Upsilon) = \nonumber \\ & \max \Biggl(1,\frac{\zeta \Upsilon_{n,k}}{\ln(2) {\min \left(\frac{P_t}{K} , \frac{I^{m}_{th}}{N^{sp}_{m}} \right)} (\mu + \eta(\Upsilon) |\hat{H}^{sp}_{m,k} (1 + \rho^2)|^2)} \Biggl).
\label{optimumConstellation2}
\end{align}
Hence, the following maximum aggregate average spectral efficiency for the spectrum-sharing system under imperfect cross-link CSI knowledge for the `average case' of estimation error can be achieved based on the optimal power, rate, and subcarrier allocation policies 
\begin{align}
& ASE^{*} = \sum_{n=1}^{N} \sum_{k=1}^{K} E_{\Upsilon | \hat{h}^{n,k}} \Bigg\{   \log_{2} \Biggl[ \max \Biggl( 1, \nonumber \\ &  \frac{\zeta \Upsilon_{n,k}}{\ln(2) \min \left(\! \frac{P_t}{K} , \frac{I^{m}_{th}}{N^{sp}_{m}} \! \right) (\mu \! + \! \eta(\Upsilon) |\hat{H}^{sp}_{m,k} (1 \! + \! \rho^2)|^2)} \Biggl) \Biggl] \varphi^{*}_{n,k}(\Upsilon) \Bigg\}
\end{align}
where the optimized cut-off SINR is expressed as: $\Upsilon^{th}_{n,k} = \frac{\ln(2) {\min \left(\frac{P_t}{K} , \frac{I^{m}_{th}}{N^{sp}_{m}} \right)} (\mu + \eta(\Upsilon) |\hat{H}^{sp}_{m,k} (1 + \rho^2)|^2)}{\zeta}$. 

\subsection{Analysis for the Worst Case of Estimation Error}  

To derive the interference constraint for the worst case scenario, we must obtain a formulation for the upper-bound of $\Delta H^{sp}_{m,k}$. Recall that $\Delta H^{sp}_{m,k}$ is a Gaussian random variable. Therefore, we proceed by bounding the channel estimation error with a certain probability. By employing the Chebyshev's inequality, for any $Y > 0$, we have 
\begin{align}
\underbrace{\mathscr{P} \biggr( |\Delta H^{sp}_{m,k} | \hat{H}^{sp}_{m,k}| \leq \Omega \biggr)}_{pr} \geq 1 - \frac{1}{Y^2}
\end{align}
where
\begin{align}
\Omega & = E_{\Delta H^{sp}_{m,k} | \hat{H}^{sp}_{m,k}}(\Delta H^{sp}_{m,k} | \hat{H}^{sp}_{m,k}) \nonumber \\ & + Y \sqrt{var(\Delta H^{sp}_{m,k} | \hat{H}^{sp}_{m,k})}).
\end{align}
With further manipulation, for a given probability of error, $pr$, the following holds
\begin{align}
\Omega & = \sqrt{\frac{var(\Delta H^{sp}_{m,k} | \hat{H}^{sp}_{m,k})}{1-pr}} + E_{\Delta H^{sp}_{m,k} | \hat{H}^{sp}_{m,k}}(\Delta H^{sp}_{m,k} | \hat{H}^{sp}_{m,k}). 
\end{align}
The interference constraint for the `worst case' scenario of estimation error is expressed as
\begin{align}
\sum_{n=1}^{N} \sum_{k=1}^{K} \varphi_{n,k}(\Upsilon) P_{n,k}(\Upsilon) |\hat{H}^{sp}_{m,k} + \Omega |^2 \leq I^{m}_{th}.
\end{align}

Utilizing one-level dual decomposition method, and by applying KKT conditions, the optimum adaptive power allocation scheme for user $n$ over subcarrier $k$ is derived as 
\begin{align}
& P^{*}_{n,k}(\Upsilon) = \nonumber \\ & \Biggr[\frac{1}{\ln(2) (\mu + \eta(\Upsilon) |\hat{H}^{sp}_{m,k} + \Omega|^2)} - \frac{{\min \left(\frac{P_t}{K} , \frac{I^{m}_{th}}{N^{sp}_{m}} \right)}}{\zeta \Upsilon_{n,k}} \Biggr]^{+}
\label{Pstar3}
\end{align}
where in the `worst case', $N^{sp}_{m} = \sum_{k=1}^{K} |\hat{H}^{sp}_{m,k} + \Omega |^2$.
To derive the optimal subcarrier allocation policy, the following maximization problem is formulated
\begin{align}
n^{*} = argmax(\Lambda(\Upsilon_{n,k})) \; , \; \forall n \in \{1,...,N\} \; , \; \forall k \in \{1,...,K\}
\label{PROBLEMSCAPOPT3}
\end{align}
where the optimal cognitive user index, $n^{*}$, can be obtained by substituting (\ref{Pstar3}) in (\ref{SCAPOPT2}) and thus solving the optimization problem in (\ref{PROBLEMSCAPOPT3}). The optimal continuous-rate solution for the constellation size of user $n$ over subcarrier $k$ is derived
\begin{align}
& M^{*}_{n,k}(\Upsilon) = \nonumber \\ & \max \Biggl(1,\frac{\zeta \Upsilon_{n,k}}{\ln(2) {\min \left(\frac{P_t}{K} , \frac{I^{m}_{th}}{N^{sp}_{m}} \right)} (\mu + \eta(\Upsilon) |\hat{H}^{sp}_{m,k} + \Omega|^2)} \Biggl).
\label{optimumConstellation2}
\end{align}
The maximum aggregate average spectral efficiency for the adaptive MQAM/OFDMA system under imperfect cross-link CSI availability for the `worst case' of estimation error with a given probability of error, $pr$, is expressed as
\begin{align}
& ASE^{*} = \sum_{n=1}^{N} \sum_{k=1}^{K} E_{\Upsilon | \hat{h}^{n,k}} \Bigg\{   \log_{2} \Biggl[ \max \Biggl( 1, \nonumber \\ &  \frac{\zeta \Upsilon_{n,k}}{\ln(2) \min \left(\frac{P_t}{K} , \frac{I^{m}_{th}}{N^{sp}_{m}} \right) (\mu + \eta(\Upsilon) |\hat{H}^{sp}_{m,k} + \Omega|^2)} \Biggl) \Biggl] \varphi^{*}_{n,k}(\Upsilon) \Bigg\}
\end{align}
where the optimized cut-off SINR is expressed as: $\Upsilon^{th}_{n,k} = \frac{\ln(2) {\min \left(\frac{P_t}{K} , \frac{I^{m}_{th}}{N^{sp}_{m}} \right)} (\mu + \eta(\Upsilon) |\hat{H}^{sp}_{m,k} + \Omega|^2)}{\zeta}$.

\section{Probabilistic Interference Constraint}

In a practical spectrum-sharing system, the collision tolerable level is confined by a maximum collision probability allowed by the licensed network. The collision tolerable level is highly dependent on the primary service type. For example, in case of real-time video streaming, a high collision probability is not desirable, however, delay-insensitive services can tolerate higher packet loss rates. In this section, we consider an underlay spectrum-sharing scenario where the primary users can tolerate a maximum collision probability $\varepsilon^{m}$, $\forall m \in \{1, ...,M\}$. We derive optimal power, rate, and subcarrier allocation algorithms for the multi-user OFDMA CR system under noisy cross-link CSI availability subject to satisfying the imposed peak aggregate power and collision probability constraints. The maximization problem can be formulated as follows. 

\textit{Problem} $\mathscr{O}_{3}$:
\begin{subequations}
\label{optimizationProb3}
\begin{gather}
\max_{\varphi_{n,k}(\Upsilon),P_{n,k}(\Upsilon)} \sum_{n=1}^{N} \sum_{k=1}^{K} E_{\Upsilon | \hat{h}^{sp}} \Big\{ \log_{2}(M_{n,k}(\Upsilon)) \varphi_{n,k}(\Upsilon) \Big\} \label{OF3a}\\ 
\text{s. t.:} \quad \text{constraints in (\ref{OF1b}), (\ref{OF1d}), (\ref{OF1be}), and (\ref{OF1e})}, \nonumber \\
\mathscr{P} \left( \sum_{n=1}^{N} \sum_{k=1}^{K} \varphi_{n,k}(\Upsilon) P_{n,k}(\Upsilon) |H^{sp}_{m,k} | \hat{H}^{sp}_{m,k}|^2 > I^{m}_{th} \right) \leq \epsilon^{m} \nonumber \\ , \forall m \in \{1, ..., M\}
\label{OF3c}
\end{gather}
\end{subequations}

We proceed by deriving a posteriori distribution of the actual cross-link given the estimated channel gains. 

\textit{Proposition 2:} 
The posterior distribution of the actual channel $H^{sp}_{m,k}$ given the estimation $\hat{H}^{sp}_{m,k}$ is a complex Gaussian random variable with respective mean and variance of
\allowdisplaybreaks{
\begin{align}
& \mu_{H^{sp}_{m,k} | \hat{H}^{sp}_{m,k}} = E_{H^{sp}_{m,k} | \hat{H}^{sp}_{m,k}}(\hat{H}^{sp}_{m,k} + \Delta H^{sp}_{m,k} | \hat{H}^{sp}_{m,k}) \nonumber \\ & \hspace*{5.5em} = E_{\hat{H}^{sp}_{m,k} | \hat{H}^{sp}_{m,k}}(\hat{H}^{sp}_{m,k} | \hat{H}^{sp}_{m,k}) \nonumber \\ & + E_{\Delta H^{sp}_{m,k} | \hat{H}^{sp}_{m,k}}(\Delta H^{sp}_{m,k} | \hat{H}^{sp}_{m,k}) = (1 + \rho^2) \hat{H}^{sp}_{m,k}  
\end{align}}
and
\begin{align}
\delta^2_{H^{sp}_{m,k} | \hat{H}^{sp}_{m,k}} & = var(\hat{H}^{sp}_{m,k} + \Delta H^{sp}_{m,k} | \hat{H}^{sp}_{m,k}) \nonumber \\ & = var(\hat{H}^{sp}_{m,k} | \hat{H}^{sp}_{m,k}) + var(\Delta H^{sp}_{m,k} | \hat{H}^{sp}_{m,k}) \nonumber \\ & + 2 cov(\Delta H^{sp}_{m,k} | \hat{H}^{sp}_{m,k},\hat{H}^{sp}_{m,k} | \hat{H}^{sp}_{m,k}) \nonumber \\ & = (1 - \rho^2) \delta^{2}_{\Delta H^{sp}_{m,k}}.
\end{align}

\begin{figure*}[ht]
\begin{minipage}[b]{0.49\linewidth}
\centering
\includegraphics[width=\textwidth]{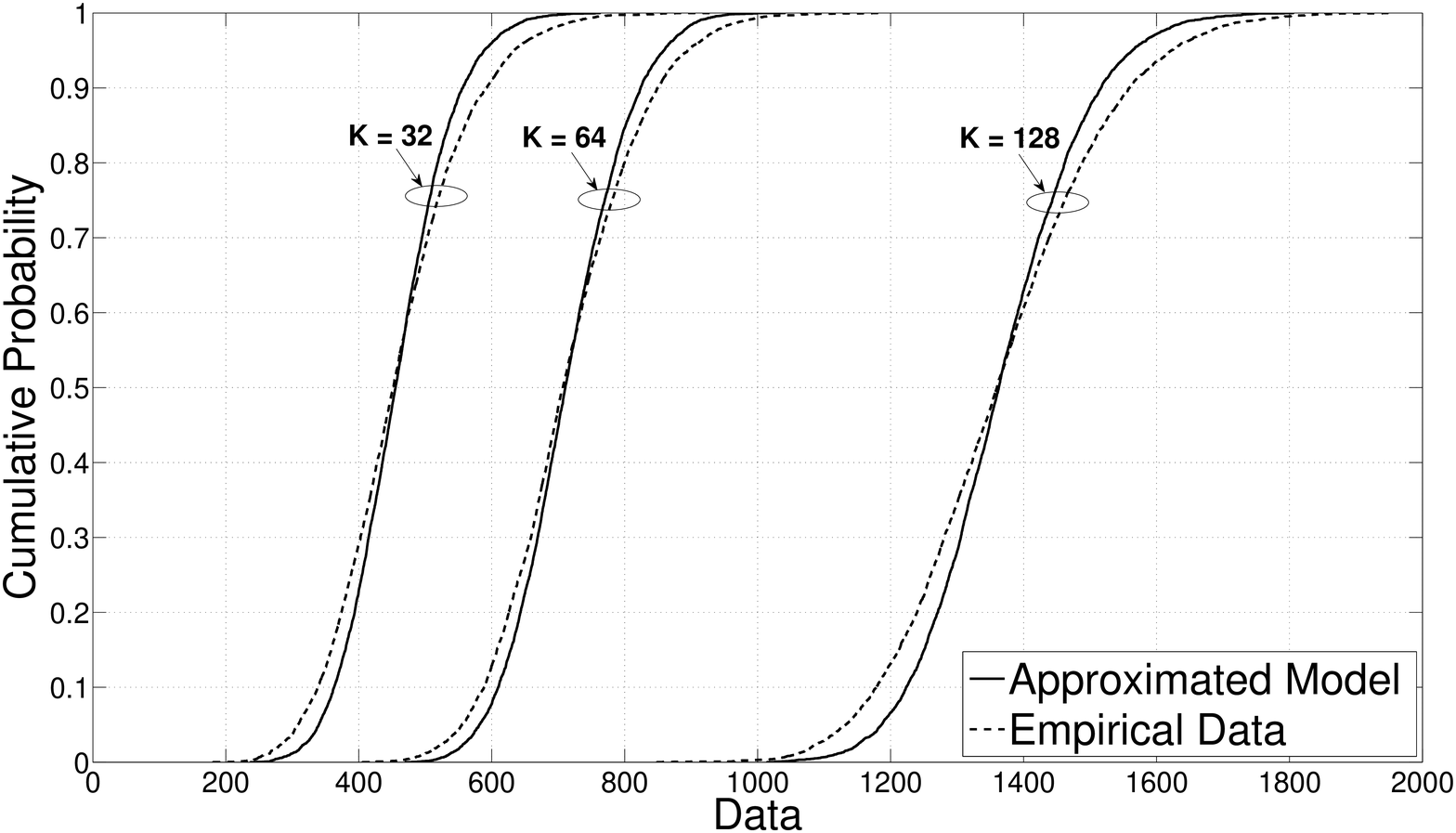}
\subcaption{$\beta^{m}_{k} \sim \text{Chi-Square}(2,2)$, $\delta^{2}_{H^{sp}_{m,k} | \hat{H}^{sp}_{m,k}}=1$.}
\label{fig:figure1}
\end{minipage}
\hfill
\begin{minipage}[b]{0.49\linewidth}
\centering
\includegraphics[width=\textwidth]{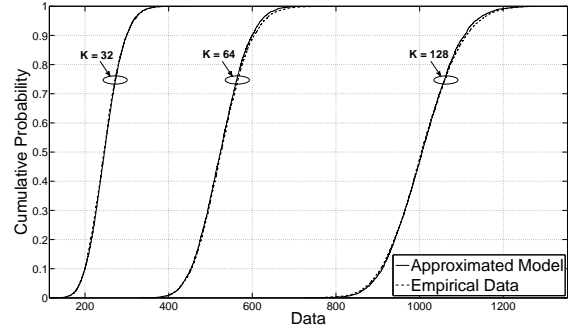}
\subcaption{$\beta^{m}_{k} \sim \text{Gamma}(2,0.5,4)$, $\delta^{2}_{H^{sp}_{m,k} | \hat{H}^{sp}_{m,k}}=0.5$.}
\label{fig:figure2}
\end{minipage}
\centering{}\caption{Approximated Model and Empirical Data cdfs, obtained from Monte-Carlo simulations.
\label{cdf-1-1}}
\end{figure*}
%

Assuming equal variance $\delta^2_{H^{sp}_{m,k} | \hat{H}^{sp}_{m,k}}$ across all users and subcarriers, the collision probability constraint in (\ref{OF3c}) can be expressed as
\begin{align}
\mathscr{P} \left(\!\delta^2_{H^{sp}_{m,k} | \hat{H}^{sp}_{m,k}} \sum_{n=1}^{N} \! \sum_{k=1}^{K} \! \varphi_{n,k}(\Upsilon) P_{n,k}(\Upsilon) |\Xi^{m}[k]|^2 > I^{m}_{th} \! \right) \! \leq \! \varepsilon^{m}
\label{CHAINOTNOT}
\end{align}
where $\Xi^{m}[k]$ is a complex Gaussian random variable with variance of one and mean of
\begin{align}
\mu_{\Xi^{m}[k]} = \abs[\Bigg]{ \frac{\mu_{H^{sp}_{m,k} | \hat{H}^{sp}_{m,k}}}{\delta_{H^{sp}_{m,k} | \hat{H}^{sp}_{m,k}}} } ^2.
\end{align}

It should be noted that in contrast to the sum of equal-weighted Chi-Square random variables in Lemma 1, (\ref{CHAINOTNOT}) includes a sum of non-equal-weighted Chi-Square random variables. 
In general, obtaining the exact distribution of the linear combination of weighted Chi-Square random variables is rather complex. Although several approximations have been proposed in the literature, e.g., \cite{springerlink:10.1007/BF02932611,springerlink:10.1007/BF02595410,1987}, most are not easy to implement. In this work, we propose a simple approximation based on the moments of $\delta^2_{H^{sp}_{m,k} | \hat{H}^{sp}_{m,k}} \sum_{n=1}^{N} \! \sum_{k=1}^{K} \! \varphi_{n,k}(\Upsilon) P_{n,k}(\Upsilon) |\Xi^{m}[k]|^2$. 
Consider the following equality
\begin{align}
\delta^2_{H^{sp}_{m,k} | \hat{H}^{sp}_{m,k}} \sum_{n=1}^{N} \! \sum_{k=1}^{K} \! \varphi_{n,k}(\Upsilon) P_{n,k}(\Upsilon) |\Xi^{m}[k]|^2 \! = \! \sum_{k=1}^{K} \! \beta^{m}_{k} |\Xi^{m}[k]|^2
\label{ProbInt}
\end{align}
where $\beta^{m}_{k} = \sum_{n=1}^{N} \delta^2_{H^{sp}_{m,k} | \hat{H}^{sp}_{m,k}} \varphi_{n,k}(\Upsilon) P_{n,k}(\Upsilon)$. 

\textit{Proposition 3:} The distribution of the sum of non-equal-weighted non-central Chi-Square random variables, i.e., $\sum_{k=1}^{K} \beta^{m}_{k}|\Xi^{m}[k]|^2$, is similar to that of a weighted non-central Chi-Square-distributed random variable $\xi \chi_{D}^{2}(\delta^{'})$, where $\delta^{'}$, $D$, and $\xi$ are respectively the non-centrality parameter, degree of freedom, and weight of the new random variable:
\begin{align}
\delta^{'} = \sum_{k=1}^{K} \mu_{\Xi^{m}[k]} \label{APsub1}\\
D = 2K \label{APsub2}\\
\xi = \frac{\sum_{k=1}^{K} \beta^{m}_{k} (2 + \mu_{\Xi^{m}[k]})}{2K + \sum_{k=1}^{K} \mu_{\Xi^{m}[k]}}. \label{APsub3}
\end{align}
To investigate the above similarity, or the accuracy of the
proposed approximation, we compare the cdf of the proposed Chi-Square distribution with that of (\ref{ProbInt}), using Monte-Carlo simulations. The results in Fig. \ref{cdf-1-1} illustrate that the approximation is accurate over a wide range of practical values for $K$ over randomly-distributed - e.g., Chi-Square or Gamma - weights $\beta^{m}_{k}$. Now (\ref{CHAINOTNOT}) can be simplified to:
\begin{align}
& \mathscr{P} \Biggr( \delta^2_{H^{sp}_{m,k} | \hat{H}^{sp}_{m,k}} \sum_{n=1}^{N} \sum_{k=1}^{K} \varphi_{n,k}(\Upsilon) P_{n,k}(\Upsilon) |\Xi^{m}[k]|^2 > I^{m}_{th} \Biggr) \nonumber \\ & \approx Pr(\xi \chi_{D}^{2}(\delta^{'}) > I^{m}_{th}).
\end{align}
According to \cite{1987}, since the non-centrality parameter is small relative to the degree of freedom, we can approximate the non-central Chi-Square distribution with a central one using the following
\begin{align}
\mathscr{P} (\xi \chi_{D}^{2}(\delta^{'}) > I^{m}_{th}) \approx \mathscr{P} (\chi_{D}^{2}(0) > \frac{I^{m}_{th}/\xi}{1 + \delta^{'}/D}).
\label{PrNCPrC}
\end{align}
The right hand side (RHS) of (\ref{PrNCPrC}) can be formulated using the upper Gamma function \cite{9761} as \\
\begin{align}
\mathscr{P}(\chi_{D}^{2}(0) > \frac{I^{m}_{th}/\xi}{1 + \delta^{'}/D}) = \frac{\Gamma(K,\frac{I^{m}_{th}/\xi}{2(1 + \delta^{'}/D)})}{\Gamma(K)}
\end{align}
where $\Gamma(.,.)$ is the upper incomplete Gamma function, and $\Gamma(.)$ is the complete Gamma function.

\textit{Proposition 4:} For all integer values $K \neq 1$, and all positive $\frac{I^{m}_{th}/\xi}{1 + \delta^{'}/D}$ - this condition is always true because, $I^{m}_{th}$, $\delta^{'}$ $\beta^{m}_{k}$, and $K$ are positive; consequently, $\xi$, $\delta^{'}$, and $D$ are also positive - the deterministic inequality 
\begin{gather}
\delta^2_{H^{sp}_{m,k} | \hat{H}^{sp}_{m,k}} \sum_{k=1}^{K} (2 + \mu_{\Xi^{m}[k]}) \sum_{n=1}^{N} \varphi_{n,k}(\Upsilon) P_{n,k}(\Upsilon) \nonumber \\ \leq \frac{K \, I^{m}_{th}}{(K!)^{1/K} \ln \left( 1 - (1 - \varepsilon^{m})^{1/K} \right)} 
\label{DetermIC}
\end{gather}
satisfies the probabilistic inequality (\ref{CHAINOTNOT}). Therefore, the constraint (\ref{CHAINOTNOT}) can be replaced by (\ref{DetermIC}).

\textit{Proof:} The proof is given in the Appendix A.

To obtain $ASE^{*}$ for the probabilistic interference constraint and `probabilistic case' of estimation error scenario, we employ the Lagrangian dual optimization method as in the previous sections, where
\begin{align}
\alpha_{k} = \delta^2_{H^{sp}_{m,k} | \hat{H}^{sp}_{m,k}} (2 + \mu_{\Xi^{m}[k]})
\end{align}
and 
\begin{align}
\overline{I^{m}_{th}} = \frac{K \, I^{m}_{th}}{(K!)^{1/K} \ln \left( 1 - (1 - \varepsilon^{m})^{1/K} \right)}.
\end{align}
Therefore, by solving the Lagrangian optimization problem the following potential optimal power allocation solution can be obtained for user $n$ over subcarrier $k$
\begin{align}
P^{*}_{n,k}(\Upsilon) = \Biggr[\frac{1}{\ln(2) (\mu + \eta(\Upsilon) \alpha_{k})} - \frac{{\min \left(\frac{P_t}{K} , \frac{\overline{I^{m}_{th}}}{\hat{N}^{sp}_{m}} \right)}}{\zeta \Upsilon_{n,k}} \Biggr]^{+} 
\label{Pstar2}
\end{align}
where $\hat{N}^{sp}_{m}$ in the `probabilistic case' is derived in Appendix A, Section C.
The optimal subcarrier allocation policy is the solution to the following problem
\begin{align}
n^{*} = argmax(\Lambda(\Upsilon_{n,k})) \; , \; \forall n \in \{1,...,N\} \; , \; \forall k \in \{1,...,K\}
\label{PROBLEMSCAPOPT4}
\end{align}
where $n^{*}$ is the optimal CRx index, and
\begin{align}
& \Lambda(\Upsilon_{n,k}) = \nonumber \\ & \frac{\frac{\zeta \Upsilon_{n,k} P^{*}_{n,k}(\Upsilon)}{{\min \left(\frac{P_t}{K} , \frac{\overline{I^{m}_{th}}}{\hat{N}^{sp}_{m}} \right)}}}{\ln(2) \left( 1 + \frac{\zeta \Upsilon_{n,k} P^{*}_{n,k}(\Upsilon)}{{\min \left(\frac{P_t}{K} , \frac{\overline{I^{m}_{th}}}{\hat{N}^{sp}_{m}} \right)}} \right)} + \frac{\ln \left( 1 + \frac{\zeta \Upsilon_{n,k} P^{*}_{n,k}(\Upsilon)}{{\min \left(\frac{P_t}{K} , \frac{\overline{I^{m}_{th}}}{\hat{N}^{sp}_{m}} \right)}} \right)}{\ln(2)}.
\label{SCAPOPT4}
\end{align}
By employing the sub-gradient method in Algorithm 1, the Lagrangian multipliers $\mu$ and $\eta(\Upsilon)$ can be updated by 
\begin{gather}
\mu^{i+1} = \mu^{i} - \tau^{i}_{1} \biggr( P_{t} - \sum_{n=1}^{N} \sum_{k=1}^{K} \varphi^{*}_{n,k}(\Upsilon) P^{*}_{n,k}(\Upsilon) \biggr) \label{SUB14} \\
\eta^{i+1}(\Upsilon) = \eta^{i}(\Upsilon) - \tau^{i}_{2} \nonumber \\ \times \left( \overline{I^{m}_{th}} \! - \! \delta^2_{H^{sp}_{m,k} | \hat{H}^{sp}_{m,k}} \sum_{k=1}^{K} (2 \! + \! \mu_{\Xi[k]}) \sum_{n=1}^{N} \varphi^{*}(\Upsilon) P^{*}_{n,k}(\Upsilon) \right). 
\label{SUB24}
\end{gather}

Subsequently, optimal expressions are derived for the constellation size and hence aggregate spectral efficiency under collision probability constraint and imperfect cross-link CSI:
\begin{align}
M^{*}_{n,k}(\Upsilon) = \max \Biggl(1,\frac{\zeta \Upsilon_{n,k}}{\ln(2) {\min \left(\frac{P_t}{K} , \frac{\overline{I^{m}_{th}}}{\hat{N}^{sp}_{m}} \right)} (\mu + \eta(\Upsilon) \alpha_{k})} \Biggl),
\label{optimumConstellation4}
\end{align}
\begin{align}
& ASE^{*} = \sum_{n=1}^{N} \sum_{k=1}^{K} E_{\Upsilon | \hat{h}^{n,k}} \Bigg\{   \log_{2} \Biggl[ \max \Biggl( 1, \nonumber \\ &  \frac{\zeta \Upsilon_{n,k}}{\ln(2) \min \left(\frac{P_t}{K} , \frac{\overline{I^{m}_{th}}}{\hat{N}^{sp}_{m}} \right) (\mu + \eta(\Upsilon) \alpha_{k})} \Biggl) \Biggl] \varphi^{*}_{n,k}(\Upsilon) \Bigg\}
\end{align}
where the optimized cut-off SINR threshold is computed by: $\Upsilon^{th}_{n,k} = \frac{\ln(2) {\min \left(\frac{P_t}{K} , \frac{\overline{I^{m}_{th}}}{\hat{N}^{sp}_{m}} \right)} (\mu + \eta(\Upsilon) |\hat{H}^{sp}_{m,k} + \Omega|^2)}{\zeta}$.

The methodologies for deriving the expressions of the cdf of the received SINR given the estimation, for different `average case', `worst case', and `probabilistic case' scenarios of estimation error, are elucidated in Appendix B. 

\section{Discussion of Results}

In this section, we examine the performance of the OFDMA CR network operating under total average transmit power and deterministic/probabilistic peak aggregate interference constraints with perfect/imperfect cross-channel estimation using the respective optimal resource allocation solutions. In the following results, perfect CSI knowledge of the cognitive user link is assumed to be available at the CTx through an error-free feedback channel. Thus, $|H^{ss}_{n,k}|$, $\forall \{n,k\}$, are drawn through a Rayleigh distribution. Further, the secondary-secondary power gain mean values, $\mu_{|H^{ss}_{n,k}|^2}$, $\forall \{n,k\}$, are taken as Uniformly-distributed random variables within 0 to 2. It should be noted that the sub-channels are assumed to be narrow-enough so that they experience frequency-flat fading. 
Interfering cross-channel values, $H^{sp}_{m,k}$, $\forall \{m,k\}$, are distributed according to a complex Gaussian distribution with mean 0.05 and variance 0.1. For the inaccurate cross-link CSI case, the channel estimation and error for all sub-channels are taken as i.i.d zero-mean Normally-distributed random variables. In addition, the AWGN power spectral density is set to -174 dBm. The total average power constraint is imposed on the system in all cases. 
Discrete-rate cases with real-valued MQAM signal constellations, i.e., $\log_{2}(M) \in \{2,4,6,8,10\}$ bits/symbol, are also considered for practical scenarios. All results correspond to the scenario with three cognitive receivers and a single primary receiver, hence, the subscript $m$ is hereafter omitted. 

\begin{figure}[t]
\includegraphics[width=.5\textwidth]{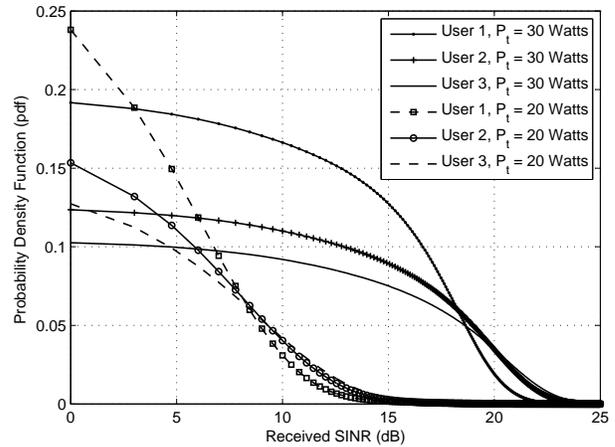} 
\caption{Probability density functions of the received SINR for OFDMA users in a given subcarrier $k$ under different average power constraint values. System parameters are: $K = 64$, $k = 16$, $I_{th} = 5$ Watts.}
\label{fig:PDFvsSINR}
\end{figure}

\begin{figure}[t]
\includegraphics[width=.5\textwidth]{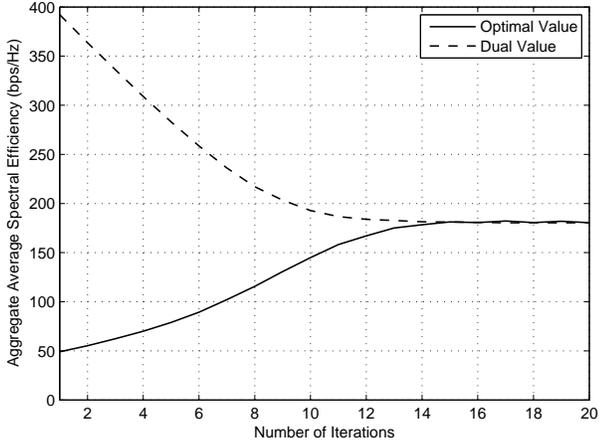} 
\caption{Optimal and dual values versus the number of iterations using the sub-gradient method. Results for the case with deterministic interference constraint and perfect cross-link CSI knowledge. System parameters are: $K = 64$, $P_t = 30$ Watts, $I_{th} = 10$ Watts, $\xi = 10^{-2}$.}
\label{fig:subgrad}
\end{figure}

The approximated probability distributions of the received SINRs for cognitive users in a randomly taken subcarrier, i.e., here $k=16$, under different total average power constraint limits $P_t$ is plotted in Fig. \ref{fig:PDFvsSINR}. For a fixed interference constraint of $I_{th} = 5$ Watts, it can be observed that the probability of higher received SINR improves as the value of $P_t$ increases. For example, for user 3, the probability of receiving $\gamma_{3,16} = 10$ dB is 54.5\% higher as the value of $P_t$ is increased from 20 to 30 Watts. 

Fig. \ref{fig:subgrad} illustrates the evolution of the optimal and dual values using the sub-gradient method over time. The results correspond to the maximum deliverable ASE for the case with deterministic interference constraint and perfect cross-link CSI knowledge. The iterative sub-gradient algorithm converges quickly and typically achieves a lower-bound at 96.5\% of the optimal value within 12 iterations. It can easily be shown that the proposed dual decomposition algorithm converges fast for different parameters of system settings. 

\begin{figure}[t]
\includegraphics[width=.5\textwidth]{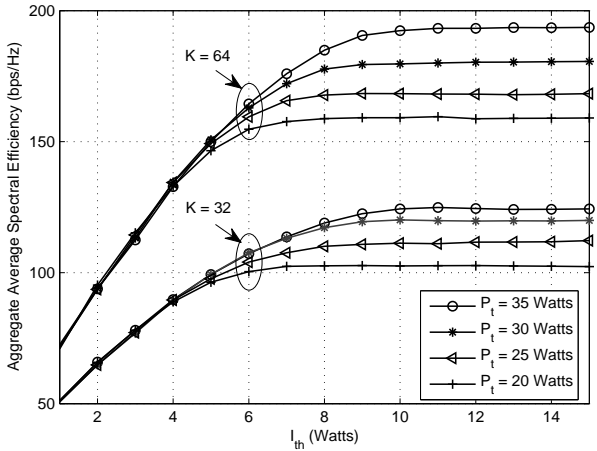} 
\caption{ASE performance versus the tolerable interference power threshold level with different values of $P_t$ and $K$. Results for the case with deterministic interference constraint and perfect cross-link CSI knowledge. System parameters are: $I_{th} = 10$ Watts, $\xi = 10^{-2}$.}
\label{fig:ASEvsPt}
\end{figure}

\begin{figure}[t]
\includegraphics[width=.5\textwidth]{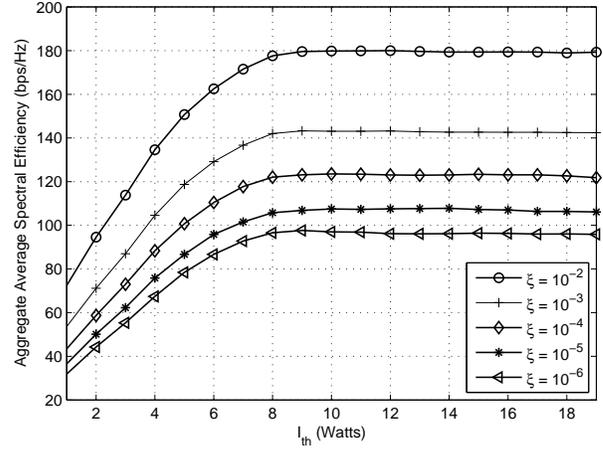} 
\caption{ASE performance using the proposed RRA algorithm versus $I_{th}$ constraint for different BER-target values. Results correspond to the case with deterministic interference constraint and perfect cross-link CSI. System parameters are: $K = 64$, $P_t = 30$ Watts.}
\label{fig:ASEvsBER}
\end{figure}

\begin{figure}[t]
\includegraphics[width=.5\textwidth]{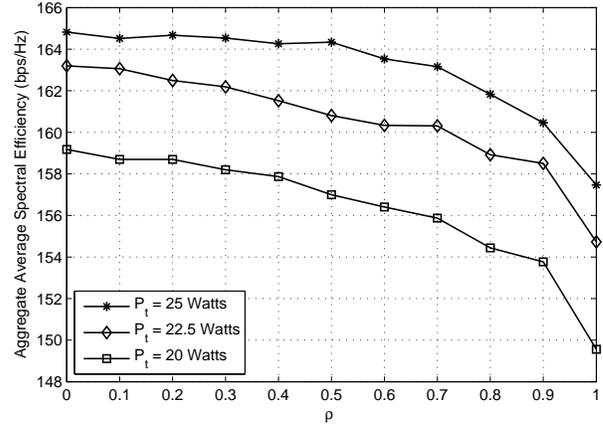} 
\caption{Achievable ASE with imperfect cross-link CSI and `average case' of estimation error against $\rho$ for different values of $P_t$. System parameters are: $K = 64$, $I_{th} = 25$ Watts, $\xi = 10^{-2}$, $\delta^{2}_{\hat{H}^{sp}_{k}} = 1$.}
\label{fig:ASEaveragecasevsRho}
\end{figure}

\begin{figure}[t]
\includegraphics[width=.5\textwidth]{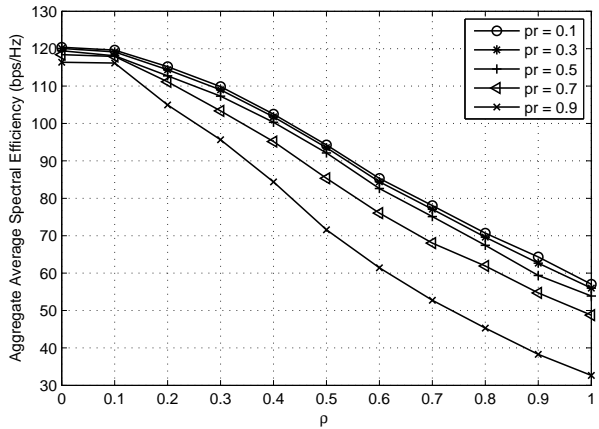} 
\caption{Achievable ASE with imperfect cross-link CSI and `worst case' of estimation error against $\rho$ with $pr$. System parameters are: $K = 64$, $P_t = 20$ Watts, $I_{th} = 5$ Watts, $\xi = 10^{-3}$, $\delta^{2}_{\hat{H}^{sp}_{k}} = 1$.}
\label{fig:ASEworstcasevsRho}
\end{figure}

\begin{figure}[t]
\includegraphics[width=.5\textwidth]{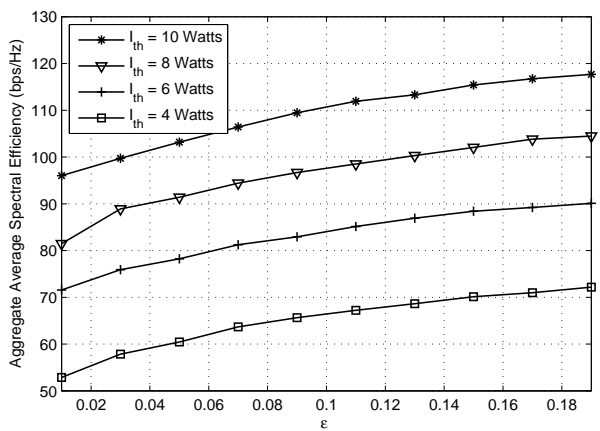} 
\caption{Achievable ASE with imperfect cross-link CSI and `probabilistic case' of estimation error against $\epsilon$ with $I_{th}$. System parameters are: $K = 64$, $P_t = 40$ Watts, $\xi = 10^{-3}$, $\rho = 0.5$, $\delta^{2}_{\hat{H}^{sp}_{k}} = 1$.}
\label{fig:ASEprobcasevsEps}
\end{figure}

\begin{figure}[t]
\includegraphics[width=.5\textwidth]{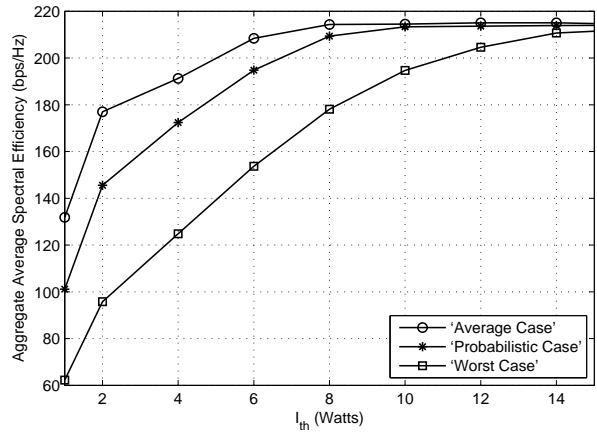} 
\caption{Performance under imperfect cross-link CSI for different cases of estimation error against $I_{th}$. System parameters are: $K = 64$, $P_t = 45$ Watts, $\xi = 10^{-2}$, $pr = 0.95$, $\rho = 0.2$, $\epsilon = 5\%$, $\delta^{2}_{\hat{H}^{sp}_{k}} = 0.1$.}
\label{fig:ASEcompareIth}
\end{figure}

Fig. \ref{fig:ASEvsPt} shows the achievable ASE of the adaptive MQAM/OFDMA CR system versus CTx-PRx interference power threshold levels under total average power and deterministic interference constraints with perfect cross-link CSI knowledge. As expected, greater ASE values are achieved for higher maximum tolerable interference since $I_{th}$ limits the cognitive users' transmit power. The improved performance however approaches a plateau in the high $I_{th}$ region as the $P_t$ threshold becomes the dominant power constraint. Note that the improved performance by increasing $I_{th}$ comes at the cost of increased probability for violating the primary users' QoS. Further, imposing a higher maximum peak average power setting enhances the achievable ASE in high $I_{th}$ region - $P_t$, for the particular values taken in this example, achieve the same ASE over small $I_{th}$ settings. Moreover, increasing the number of subcarriers results in higher attainable performance.    

Achievable ASE performance under different maximum tolerable interference thresholds for respective values of BER-target with perfect cross-link CSI availability is shown in Fig. \ref{fig:ASEvsBER}. It can be seen that the system performance is improved under less stringent QoS constraints. For example, a 26.9\% gain in ASE performance is achieved by imposing $\xi = 10^{-2}$ in comparison to $\xi = 10^{-3}$. However, the gap in performance becomes less significant for lower BER-target regimes. 

System performance with noisy cross-link CSI and `average case' of estimation error versus the correlation factor between estimation and error variables $\rho$ is depicted in Fig. \ref{fig:ASEaveragecasevsRho}. It can be seen that a higher correlation factor increases the maximum likelihood between true and estimated interfering channels, hence, the probability of violating the interference constraint on average is improved and in turn a lower ASE for the cognitive system is realized.  
Further, the achievable ASE with imperfect cross-link CSI knowledge and `worst case' of estimation error against $\rho$ for different probabilities of channel estimation error bound $pr$ is studied in Fig. \ref{fig:ASEworstcasevsRho}. Apart from the effect of $\rho$ on the performance, higher values of $pr$ increase the robustness of the interference management scheme but come at the cost of lower achievable spectral efficiencies. The results indicate that the improved ASE performance by decreasing $pr$ in the lower half region (i.e., $pr \leq 0.5$) is not significant yet it may cause critical interference to the primary service operation. For example, given $\rho = 0.5$, varying the value of $pr$ from 0.5 to 0.1 results in a 40\% increase in the probability of error bound violation but only provides an effective gain of 2.3\% in the cognitive system performance.      

The achievable performance with imperfect cross-channel information and `probabilistic case' of estimation error versus the collision probability $\epsilon$ with respective $I_{th}$ values is illustrated in Fig. \ref{fig:ASEprobcasevsEps}. Increasing the maximum probability of violating the interference constraint set by the a regulatory authority significantly improves the spectral efficiency of the cognitive network. The tradeoff is however the degradation of the primary service operation which is deemed highly undesirable in practical scenarios.   

System performance with noisy cross-link CSI for different `average case', `worst case', and `probabilistic case' of estimation error is demonstrated in Fig. \ref{fig:ASEcompareIth}. The results show that the `probabilistic case' with $5\%$ collision probability outperforms the achievable ASE under the `worst case' scenario with an error bound of $pr = 0.5$. For example, given $I_{th} = 6$ Watts, the `probabilistic case' achieves a 26.7\% gain in ASE over the `worst case'. Further, employing the `average case' provides higher spectral efficiencies. For instance, a 7.0\% increase in performance in achieved utilizing the `average case' over the `probabilistic case'. For high values of $I_{th}$, the total average power constraint becomes the dominant limit and therefore the performance under different cases of estimation error eventually converge. Note that the `average case' controls the interference based on the average error estimation, therefore, it cannot mitigate the potential instantaneous interference violations. On the other hand, implementing the `worst case' can guarantee that the interference constraints are obeyed at any given time, thus, preserving the primary users' QoS. The proposed `probabilistic case' of estimation error provides an optimal trade-off  between the achievable performance of cognitive system and managing the QoS of primary users. In particular, the `probabilistic case' is advantageous in terms of performance and flexibility over the conventional `average case' and `worst case' scenarios. 

\section{conclusions}

In this paper, we have studied the spectral efficiency performance of adaptive MQAM/OFDMA underlay CR networks with certain/uncertain interfering channel information. We derived novel RRA algorithms to enhance the overall cognitive system performance subject to satisfying total average power and peak aggregate interference constraints. The proposed framework considers  both cases of perfect and imperfect cross-link CSI knowledge at the cognitive transmitter. In the latter, different `average case', `worst case', and `probabilistic case' scenarios of channel estimation error were modeled and analysed. To compute the aggregate average spectral efficiency, we developed unique approximated distributions of the received SINR for given users over different sub-channels in the respective cases under consideration. Through simulation results we studied the achievable performance of the cognitive system using our proposed RRA algorithms. By adapting the power, rate, and subcarrier allocation policies to the time-varying secondary-secondary fading channels and secondary-primary interfering channels, a significant gain in the spectral efficiency performance of the cognitive system can be realized, whilst controlling the interference on the primary service receivers. Furthermore, the impact of parameters uncertainty on overall system performance was investigated. In particular, simulation results were provided for different cases of error estimation. It was understood that the `average case' results in higher cognitive system performance, however, comes at the cost of potential instantaneous interference violations. Subsequently, the `worst case' can guarantee that the power constraints are obeyed at all times, yet it does not result in desirable cognitive performance. In contrast, the proposed `probabilistic case' in this paper, which was derived as a low complexity deterministic constraint, provided an optimal trade-off between the achievable performance of the cognitive network and preserving the QoS of the primary users. In summary, the `probabilistic case' can replace the conventional `average case' and `worst case' scenarios in practical situations as a result of enhanced performance and flexibility.

\appendices
\section{Proof of Proposition 4}

For a Chi-Square random variable $\chi_{D}^{2}$, with a degree of freedom $2K$, the probability
\begin{equation}
\mathscr{P} (\xi\chi_{D}^{2}(\delta^{'}) > I_{th}^{m})\approx \mathscr{P} (\xi\chi_{D}^{2}(0) > \frac{I_{th}^{m}/\xi}{1+\nicefrac{\delta^{'}}{D}}),
\end{equation}
can be formulated using the upper gamma function \cite{Alzer97onsome} as
\begin{equation}
Pr(\chi_{D}^{2}(0) > \frac{I_{th}^{m}/\xi}{1+\nicefrac{\delta^{'}}{D}})=\frac{\Gamma(K\,,\,\frac{I_{th}^{m}}{2 \xi(1+\nicefrac{\delta^{'}}{D})})}{\Gamma(K)}.
\end{equation}
By defining
\begin{equation}
x^{y}=\frac{I_{th}^{m}}{2 \xi (1+\nicefrac{\delta^{'}}{D})},\label{eq:meghdar x}
\end{equation}
and using the results from \cite{Alzer97onsome}, we have
\begin{equation}
\Gamma(K\,,\, x^{y})=y\int_{x}^{\infty}e^{-t^{y}}dt,
\end{equation}
where $y=1/K$. Given $1/(K\Gamma(K))=1/\Gamma(K+1)$,
and using the corollary of \cite{Alzer97onsome}, we can derive the following equalities
\begin{equation}
\frac{\Gamma(K\,,\, x^{y})}{\Gamma(K)}=\frac{y\int_{x}^{\infty}e^{-t^{y}}dt}{\Gamma(K)}=\frac{\int_{x}^{\infty}e^{-t^{y}}dt}{\Gamma(K+1)}.\label{eq:gamma2}
\end{equation}
The upper-bound of (\ref{eq:gamma2}) can be expressed as
\begin{equation}
\frac{\Gamma(K\,,\, x^{y})}{\Gamma(K)}=\frac{\int_{x}^{\infty}e^{-t^{y}}dt}{\Gamma(K+1)}\leq1-[1-e^{-\vartheta x^{y}}]^{1/y}.\label{eq:gamma}
\end{equation}
The expression in (\ref{eq:gamma}) is valid for all positive $x$, if and only if,
\begin{equation}
0\leq\vartheta\leq \min\{1,[\Gamma(K+1)]^{-1/K}\}.
\end{equation}
Thus, for all $K\geq1$,
\begin{equation}
\vartheta=(\Gamma(K+1))^{-1/K}.\label{eq:tawan}
\end{equation}
Subsequently, from (\ref{CHAINOTNOT}), (\ref{eq:gamma}), and (\ref{eq:tawan}),
\begin{equation}
\frac{\Gamma(K\,,\, x^{y})}{\Gamma(K)}\leq\varepsilon^{m} \label{eq:tawan-1}
\end{equation}
and by replacing (\ref{eq:gamma2}) in (\ref{eq:tawan-1}), we have
\begin{equation}
1-[1-e^{-\vartheta x^{y}}]^{1/y}\leq\varepsilon^{m}.\label{eq:tawan-2}
\end{equation}
With further manipulation, it can be shown that
\begin{equation}
x^{y}=-\frac{\ln(1-\sqrt[K]{1-\varepsilon^{m}})}{(\Gamma(K+1))^{-1/K}}.\label{eq:tawan-3}
\end{equation}
Replacing (\ref{eq:meghdar x}) in (\ref{eq:tawan-3}) we have: 
\begin{equation}
\frac{I_{th}^{m}}{2 \xi (1+\nicefrac{\delta^{'}}{D})}\leq-\frac{\ln(1-\sqrt[K]{1-\varepsilon^{m}})}{(\Gamma(K+1))^{-1/K}},\label{eq:tawan-3-1}
\end{equation}
Finally, by replacing (\ref{APsub1}), (\ref{APsub2}),
and (\ref{APsub3}) in (\ref{eq:tawan-3-1}), we have: 
\begin{gather}
\delta^2_{H^{sp}_{m,k} | \hat{H}^{sp}_{m,k}} \sum_{k=1}^{K} (2 + \mu_{\Xi^{m}[k]}) \sum_{n=1}^{N} \varphi_{n,k}(\Upsilon) P_{n,k}(\Upsilon) \nonumber \\ \leq \frac{K \, I^{m}_{th}}{(K!)^{1/K} \ln \left( 1 - (1 - \varepsilon^{m})^{1/K} \right)}. 
\end{gather}

\section{Received SINR cdf Derivation}

To derive the cdf of the received SINR given the estimation, $F_{\gamma_{n,k} | \hat{H}^{sp}_{m,k}}(\gamma_{n,k} | \hat{H}^{sp}_{m,k})$, for different `average case', `worst case', and `probabilistic case', scenarios.

\subsection{`Average Case'}

The random variable $H^{sp}_{m,k}|\hat{H}^{sp}_{m,k}$, given the `average case' of estimation error, is a complex Gaussian random variable with mean zero and variance $\delta^2_{\hat{H}^{sp}_{m,k}} (1 + \rho^2)^2$. Using Lemma 1, $N^{sp}_{m} = \sum_{k=1}^{K} |H^{sp}_{m,k}|\hat{H}^{sp}_{m,k}|^2$ can be approximated by 
\begin{align}
N^{sp}_{m} = \sum_{k=1}^{K} |H^{sp}_{m,k}|^2 \thicksim N \biggr( \mu_{N^{sp}_{m}} , \delta^{2}_{N^{sp}_{m}} \biggr)
\label{NCCtoN}
\end{align}
where $\mu_{N^{sp}_{m}} = 2K \delta^2_{\hat{H}^{sp}_{m,k}} (1 + \rho^2)^2$ and $\delta^{2}_{N^{sp}_{m}} = 4K \delta^4_{\hat{H}^{sp}_{m,k}} (1 + \rho^2)^4$. 
By replacing these new parameters in (\ref{CDFofgamma}), $F_{\gamma_{n,k} | \hat{H}^{sp}_{m,k}}(\gamma_{n,k} | \hat{H}^{sp}_{m,k})$ can be obtained under the `average case' of estimation error. 
 
\subsection{`Worst Case'} 

To derive the distribution of the received SINR given the estimation, for the `worst case' of estimation error, we invoke Lemma 1. In this case, the random variable $H^{sp}_{m,k}|\hat{H}^{sp}_{m,k}$ is a complex Gaussian random variable with mean $\sqrt{\frac{\delta^2_{\Delta H^{sp}_{m,k}} (1 - \rho^2)}{1 - pr}}$ and variance $\delta^2_{\hat{H}^{sp}_{m,k}} (1 + \rho^2)^2$. Hence, the random variable $N^{sp}_{m} = \sum_{k=1}^{K} |H^{sp}_{m,k}|\hat{H}^{sp}_{m,k}|^2$ can be estimated as 
\begin{align}
N^{sp}_{m} = \sum_{k=1}^{K} |H^{sp}_{m,k}|\hat{H}^{sp}_{m,k}|^2 \thicksim N \biggr( \mu_{N^{sp}_{m}} , \delta^{2}_{N^{sp}_{m}} \biggr)
\label{NCCtoN}
\end{align}
where $\mu_{N^{sp}_{m}} = \delta^2_{\hat{H}^{sp}_{m,k}} (1 + \rho^2)^2 \Big[ 2K + \mu^{'} \Big]$ and $\delta^{2}_{N^{sp}_{m}} = \delta^4_{\hat{H}^{sp}_{m,k}} (1 + \rho^2)^4 \Big[ 4K + 4 \mu^{'} \Big]$ and $\mu^{'} = \sum_{k=1}^{K} | \delta^2_{\Delta H^{sp}_{m,k}} (1 - \rho^2) / (1 - pr) \delta^2_{\hat{H}^{sp}_{m,k}} (1 + \rho^2)^2 |$. Using the parameters of $N^{sp}_{m}$ in (\ref{CDFofgamma}) yields the distribution of the received SINR given the estimation for this case.

\subsection{`Probabilistic Case'}

For the `probabilistic case' of estimation error, invoking Lemma 1 and using  (\ref{DetermIC}), 
\begin{align}
\hat{N}^{m}_{sp} & = \sum_{k=1}^{K} \delta^2_{H^{sp}_{m,k} | \hat{H}^{sp}_{m,k}} \left( 2 + \abs[\Bigg]{ \frac{\mu_{H^{sp}_{m,k} | \hat{H}^{sp}_{m,k}}}{\delta_{H^{sp}_{m,k} | \hat{H}^{sp}_{m,k}}} } ^2 \right) \nonumber \\ & = \sum_{k=1}^{K} (1 - \rho^2) \delta^2_{\Delta H^{sp}_{m,k}} \left(2 + \abs[\Bigg]{\frac{(1 + \rho^2) \hat{H}^{sp}_{m,k}}{\sqrt{(1 - \rho^2)} \delta_{\Delta H^{sp}_{m,k}}}}^2 \right)
\end{align}
is a Normally-distributed random variable with mean $\mu_{\hat{N}^{m}_{sp}} = 2K \delta^2_{\hat{H}^{sp}_{m,k}} \left( 1 + \rho^2 \right)^2 + 2 K \left( 1 - \rho^2 \right)^2 \delta^2_{\Delta H^{sp}_{m,k}}$ and variance $\delta^{2}_{\hat{N}^{m}_{sp}} = 4K \delta^4_{\hat{H}^{sp}_{m,k}} \left( 1 + \rho^2 \right)^4$. 
Hence, by replacing $\hat{N}^{m}_{sp}$ and $\overline{I^{m}_{th}}$ with $N^{sp}_{m}$ and $I^{m}_{th}$, respectively, and applying the analysis in (11)-(20), $F_{\gamma_{n,k} | \hat{H}^{sp}_{m,k}}(\gamma_{n,k} | \hat{H}^{sp}_{m,k})$ can be developed for the collision probability constraint with `probabilistic case' of estimation error. 

\bibliographystyle{IEEEtran}
{\footnotesize
\bibliography{IEEEabrv,myref}}

\end{document}